\documentclass[12pt,onecolumn,final]{IEEEtran}

\usepackage[nomarkers,nolists,noheads]{endfloat}
\usepackage{cite}
\ifCLASSINFOpdf
  \usepackage{graphicx,epstopdf}
    \DeclareGraphicsExtensions{.eps}
    \fi
\usepackage[cmex10]{amsmath}
\usepackage{algorithmic}
\usepackage[tight,footnotesize]{subfigure}
\usepackage{amssymb}
\begin{document}
\title{Large System Analysis of Linear Precoding in MISO Broadcast Channels with Confidential Messages}
\author{\normalsize\authorblockN{{Giovanni~Geraci$^{1,2}$,~Romain~Couillet$^3$,~Jinhong~Yuan$^1$,~M\'{e}rouane~Debbah$^4$~and~Iain~B.~Collings$^2$}}\\
\small\authorblockA{$^1$School of Electrical Engineering \& Telecommunications,
The University of New South Wales, \textsc{Australia} }\\
\authorblockA{$^2$Wireless and Networking Technologies Laboratory, CSIRO ICT Centre, Sydney, \textsc{Australia}}\\
\authorblockA{$^3$Dept. of Telecommunications and $^4$Alcatel-Lucent Chair on Flexible Radio,
Sup\'{e}lec, Gif-sur-Yvette, \textsc{France}}
}
\maketitle

\begin{abstract}
In this paper, we study the performance of regularized channel inversion (RCI) precoding in large MISO broadcast channels with confidential messages (BCC). We obtain a deterministic approximation for the achievable secrecy sum-rate which is almost surely exact as the number of transmit antennas $M$ and the number of users $K$ grow to infinity in a fixed ratio $\beta=K/M$. We derive the optimal regularization parameter $\xi$ and the optimal network load $\beta$ that maximize the per-antenna secrecy sum-rate. We then propose a linear precoder based on RCI and power reduction (RCI-PR) that significantly increases the high-SNR secrecy sum-rate for $1<\beta<2$. Our proposed precoder achieves a per-user secrecy rate which has the same high-SNR scaling factor as both the following upper bounds: (i) the rate of the optimum RCI precoder without secrecy requirements, and (ii) the secrecy capacity of a single-user system without interference. Furthermore, we obtain a deterministic approximation for the secrecy sum-rate achievable by RCI precoding in the presence of channel state information (CSI) error. We also analyze the performance of our proposed RCI-PR precoder with CSI error, and we determine how the error must scale with the SNR in order to maintain a given rate gap to the case with perfect CSI.
\end{abstract}

\IEEEpeerreviewmaketitle
\thispagestyle{empty}

\newpage
\section{Introduction}
\setcounter{page}{1}

Wireless networks are becoming more and more pervasive, with users relying on them to transmit sensitive data. Due to the broadcast nature of the physical medium, every node in the network is a potential eavesdropper, and securing the transmitted information is critical. Security of wireless communications has traditionally been ensured by network layer key-based cryptography. However, these schemes may not be suitable in the case of large dynamic wireless networks, since they raise issues like key distribution and management (for symmetric cryptosystems) and high computational complexity (for asymmetric cryptosystems). Moreover, these schemes are potentially vulnerable, because they rely on the unproven assumption that certain mathematical functions are hard to invert \cite{Mukherjee10Survey}. To provide an additional level of protection and to achieve perfect secrecy, methods exploiting the randomness inherent in noisy channels, known as physical layer security, have been proposed \cite{Wyner75,Csiszar78}.

Physical layer security has recently become an active area of research \cite{LiangBook,LiuBook}. The maximum rate at which a message can be reliably transmitted to an intended user while the rate of information leakage at an eavesdropper vanishes asymptotically with the code length, denoted as the secrecy capacity, has been studied for several network topologies. These include multi-antenna wiretap channels \cite{Khisti10I} and multi-receiver wiretap channels \cite{Ekrem11
}. Techniques like artificial noise and adaptive encoding have been proposed for the case when the eavesdropper's channel is not known by the transmitter \cite{Goel08,Zhou10,Mukherjee11Robust,
Zhou11}. The secrecy capacity of a two-user broadcast channel with confidential messages (BCC) has also been studied in \cite{Liu10}, where the intended users can act maliciously as eavesdroppers. For a larger BCC with any number of malicious users, practical linear precoding schemes have been proposed \cite{GeraciISWCS11}. Although suboptimal, linear precoding can control the amount of interference and information leakage between the users of a BCC, thus achieving secrecy with low-complexity implementation \cite{Geraci12,GeraciWCNC12}.

In this paper, we propose a linear precoder based on regularized channel inversion (RCI) for the multiple-input single-output (MISO) BCC. In the MISO BCC, a base station (BS) equipped with $M$ antennas simultaneously transmits $K$ independent confidential messages to $K$ spatially dispersed single-antenna users that potentially eavesdrop on each other. Under this system setup, we carry out a large-system analysis assuming that both $M$ and $K$ grow large, while their ratio $\beta = K/M$ is fixed. This paper directly extends some of the analysis in \cite{Nguyen09,Wagner12} by requiring the transmitted messages to be kept confidential. Furthermore, this paper generalizes the results provided in \cite{Geraci12}, where the special case of $\beta=1$ with perfect channel state information (CSI) was considered.  

Our main contributions can be summarized as follows.
\begin{itemize}
\item We obtain a deterministic equivalent for the large-system secrecy sum-rate achievable by the RCI precoder in a MISO BCC. We then derive the optimal regularization parameter $\xi$ that maximizes the secrecy sum-rate. Numerical results confirm that our analysis is accurate even for finite $M$. \item We derive a closed-form approximation for the optimal network load $\beta$ that maximizes the per-antenna secrecy sum-rate. We find that for $\beta >1$ the RCI precoder performs poorly in the high-SNR regime. Therefore, we propose a linear precoder based on RCI and power reduction (RCI-PR) that significantly increases the high-SNR secrecy sum-rate for $1<\beta<2$.
\item We compare the performance of our proposed RCI-PR precoder to: (i) the sum-rate achieved by an optimized RCI precoder without secrecy requirements and (ii) the secrecy capacity of a single-user system. The gaps with these two upper bounds represent the loss caused by the presence of: (i) secrecy requirements and (ii) interference due to multiple users, respectively. Both analysis and simulations show that the rate of the proposed precoder has the same high-SNR scaling factor as the two upper bounds.
\item We obtain a deterministic equivalent for the secrecy sum-rate achievable by RCI precoding in the presence of CSI error. We then analyze the performance of our proposed RCI-PR precoder and determine how the CSI estimation error must scale with the SNR, to maintain a given rate gap to the case with perfect CSI. We find that in large frequency division duplex (FDD) systems, under random vector quantization (RVQ), $B \approx \frac{M-1}{3} \rho_{\textrm{dB}} - \left( M-1 \right) \left[ \log_2 \left( \sqrt{4b-3} - 1 \right)-1\right]$ feedback bits per user are sufficient to maintain a gap of $\log_2 b$ bps/Hz at high SNR.
\end{itemize}

The remainder of the paper is organized as follows. Section II presents the system model and a secrecy sum-rate achievable by RCI precoding. In Section III, we derive a deterministic equivalent for this secrecy sum-rate, as well as the optimal regularization parameter $\xi$ and network load $\beta$. In Section IV, we propose the RCI-PR precoder, and we study the secrecy sum-rate that it can achieve. Section V compares the performance of the proposed precoder to two upper bounds, obtained (i) in the absence of secrecy requirements and (ii) in the absence of interference. In Section VI we study the case when only imperfect CSI is available at the transmitter. Section VII concludes the paper.

Throughout the paper we use the following notation: bold uppercase (lowercase) letters denote matrices (column vectors); $(\cdot)^{T}$, $(\cdot)^H$, $\mathrm{tr}\{\cdot\}$, $\| \cdot \|$, and $\mathrm{E}[\cdot]$ denote transpose, conjugate transpose, trace, Euclidean norm, and expectation, respectively; $\mathcal{CN}(\mu,\sigma^2)$ denotes circularly symmetric complex-Gaussian distribution with mean $\mu$ and variance $\sigma^2$, and we use the notation $[ \cdot ]^+ \triangleq \max(\cdot,0)$.
\section{System Model}

In this section we introduce the MISO BCC and the secrecy sum-rate achievable by RCI precoding.

\subsection{MISO Broadcast Channel with Confidential Messages}

We consider the downlink of a narrowband MISO BCC, consisting of a base station with $M$ antennas which simultaneously transmits $K$ independent confidential messages to $K$ spatially dispersed single-antenna users. In this model, transmission takes place over a block fading channel, and the transmitted signal is $\mathbf{x} = \left[x_1,\ldots,x_M \right]^{T} \in \mathbb{C}^{M \times 1}$. The received signal at user $k$ is given by
\begin{equation}
y_k=\sum_{j=1}^{M} h_{k,j}x_{j}+n_{k}
\label{eqn:MIMO_scalar}
\end{equation}
where $h_{k,j} \sim \mathcal{CN}(0,1)$ is the i.i.d. channel between the $j$-th transmit antenna element and the $k$-th user, and $n_{k} \sim \mathcal{CN}(0,\sigma^2)$ is the noise seen at the $k$-th receiver. 
The corresponding vector equation is
\begin{equation}
\mathbf{y}=\mathbf{Hx}+\mathbf{n}
\label{eqn:MIMO_vector}
\end{equation}
where $\mathbf{H} = \left[\mathbf{h}_1,\ldots,\mathbf{h}_K \right]^{\dagger}$ 
is the $K \times M$ channel matrix. We assume that E $[ \mathbf{nn}^{H} ] =\sigma^{2} \mathbf{I}$, define $\rho \triangleq 1/ \sigma ^2$, and impose the long-term power constraint E$[ \left\| \mathbf{x} \right\|^{2} ] =1$.

It is required that the BS securely transmits each confidential message $u_k$, ensuring that the unintended users receive no information. This is performed at the secrecy rate $R_{s,k}$, defined as follows. Let $\mathrm{Pr}(\mathcal{E}_n)$ be the probability of error at the $k$-th intended user, $m$ be a confidential message with entropy $H(m)$, $\mathbf{y}_k^n$ be the vector of all signals received by the eavesdroppers, and $H(m | \mathbf{y}_k^n)$ be the corresponding equivocation. Then a (weak) secrecy rate $R_{s,k}$ for the intended user is achievable if there exists a sequence of $(2^{nR_{s,k}},n)$ codes such that $\mathrm{Pr}(\mathcal{E}_n) \rightarrow 0$ and $\frac{1}{n}H(m | \mathbf{y}_k^n) \leq \frac{1}{n} H(m) - \varepsilon_n$ for some sequence $\varepsilon_n$ approaching zero as $n \rightarrow \infty$ \cite{Khisti10I}.

In general, the behavior of the users cannot be determined by the BS. 
As a worst-case scenario, in our system we assume that for each intended receiver $k$ the remaining $K-1$ users can cooperate to jointly eavesdrop on the message $u_k$.
For each user $k$, the alliance of the $K-1$ cooperating eavesdroppers is equivalent to a single eavesdropper with $K-1$ receive antennas, which is denoted by $\widetilde{k}$.

\subsection{Regularized Channel Inversion Precoding}

In this paper, we consider linear precoding for the MISO BCC. Although suboptimal, linear precoding schemes are of particular interest because of their low-complexity implementations \cite{Spencer04Magazine,Li10a}. Linear precoding can control the amount of interference to maintain a high sum-rate in the broadcast channel \cite{Spencer04,YooJSAC06,Peel05,Joham05,Sun10,Jin08}.
In the MISO BCC, linear precoding can be employed to control the amount of interference and information leakage to the unintended receivers introduced by the transmission of each confidential message \cite{GeraciISWCS11,Geraci12,GeraciWCNC12}. We are interested primarily in the RCI precoder, because it gives a better performance than the plain Channel Inversion precoder, particularly at low SNR \cite{Peel05}.

In RCI precoding, the transmitted vector $\mathbf{x}$ is obtained at the BS by performing a linear processing on the vector of confidential messages $\mathbf{u} = \left[u_1,\ldots,u_K \right]^{T}$, whose entries are chosen independently, satisfying E$[ \left|u_k\right|^2 ] =1$. We assume homogeneous users, i.e. each user experiences the same received signal power on average, thus the model assumes that their distances from the transmitter are the same. The transmitted signal $\mathbf{x}$ after RCI precoding can be written as $\mathbf{x} = \mathbf{Wu}$, 
where $\mathbf{W} = \left[\mathbf{w}_1,\ldots,\mathbf{w}_K \right]$ is the $M \times K$ RCI precoding matrix \cite{Peel05,Joham05}, given by
\begin{equation}
\mathbf{W} = \frac{1}{\sqrt{\gamma}} \mathbf{H}^H \left( \mathbf{H H}^H + M \xi \mathbf{I}_K \right) ^{-1} = \frac{1}{\sqrt{\gamma}} \left( \mathbf{H}^H \mathbf{H} + M \xi \mathbf{I}_M \right) ^{-1} \mathbf{H}^H
\label{eqn:RCI_precoder}
\end{equation}
and $\gamma = \textrm{tr} \left\{ \mathbf{H}^H \mathbf{H} \left( \mathbf{H}^H \mathbf{H} + M \xi \mathbf{I}_M \right) ^{-2} \right\}$
is a long-term power normalization constant. The function of the real regularization parameter $\xi$ is to achieve a tradeoff between the signal power at the intended user and the interference and information leakage at the other unintended users for each message. Note that unlike \cite{Nguyen09,Wagner12,Geraci12,GeraciWCNC12}, we do not confine ourselves to nonnegative regularization parameters.

\subsection{Achievable Secrecy Sum-Rates}

A secrecy sum-rate achievable by RCI precoding in the MISO BCC was obtained in \cite{Geraci12} by considering that each user $k$, along with its own eavesdropper $\tilde{k}$ and the transmitter, forms an equivalent multi-input, single-output, multi-eavesdropper (MISOME) wiretap channel \cite{Khisti10I}. The transmitter, the intended receiver, and the eavesdropper of each MISOME wiretap channel are equipped with $M$, $1$, and $K-1$ virtual antennas, respectively. Due to the assumption of cooperating malicious users, interference cancellation can be performed at 
$\tilde{k}$, which does not see any undesired signal term apart from the received noise. As a result, a secrecy sum-rate achievable by RCI precoding is given by \cite{Geraci12}
\begin{equation}
R_{s} = \sum_{k=1}^{K} R_{s,k} \mathrm{ } = \sum_{k=1}^{K} \left[ \log_2 \Big( 1 + \mathrm{SINR}_{k} \Big) - \log_2 \Big( 1 + \mathrm{SINR}_{\widetilde{k}} \Big) \right]^+,
\label{eqn:Rs}
\end{equation}
where $\mathrm{SINR}_{k}$ and $\mathrm{SINR}_{\tilde{k}}$ are the signal-to-interference-plus-noise ratios for the message $u_k$ at the intended receiver $k$ and the eavesdropper $\widetilde{k}$, respectively, given by
\begin{equation}
\mathrm{SINR}_{k} = \frac {\rho \left| \mathbf{h}_k^H \mathbf{w}_k \right| ^2} {1 + \rho \sum_{j \neq k} {\left| \mathbf{h}_k^H \mathbf{w}_j \right| ^2} } \quad \textrm{and} \quad
\mathrm{SINR}_{\widetilde{k}} = \rho \left\| \mathbf{H}_k \mathbf{w}_{k} \right\| ^2,
\label{eqn:SINR}
\end{equation}
where $\mathbf{H}_k$ is the matrix obtained from $\mathbf{H}$ by removing the $k$-th row.
\section{Large System Analysis}
In this section, we study the secrecy sum-rate of the RCI precoder in the large-system regime, where both the number of receivers $K$ and the number of transmit antennas $M$ approach infinity, with their ratio $\beta = K/M$ being held constant. We then derive the optimal regularization parameter $\xi$ that maximizes the secrecy sum-rate and a closed-form approximation for the optimal network load $\beta$.

\subsection{Deterministic Equivalent of the Secrecy Sum-Rate with RCI Precoding}
In the following we provide a deterministic approximation of the per-antenna secrecy sum-rate, which is almost surely exact as $M \rightarrow \infty$. To obtain such deterministic approximation, we need to ensure that the minimum eigenvalue of $\left(\frac{1}{M}\mathbf{HH}^H+\xi\mathbf{I}\right)$ is bounded away from zero for all large $M$, almost surely. Let $C>0$, $\epsilon>0$, we define the set $\mathcal{D}_M = \mathbb{R} \backslash \left\{ \left[-\left(1+\sqrt{\beta}\right)^2 - \frac{C}{M^{\frac{1}{2}-\epsilon}}, -\left(1-\sqrt{\beta}\right)^2 + \frac{C}{M^{\frac{1}{2}-\epsilon}} \right] \right\}$.
\newtheorem{Theorem}{Theorem}
\begin{Theorem}
Let $\rho>0$ and $\beta>0$. Let $R_s$ be the secrecy sum-rate achievable by RCI precoding defined in (\ref{eqn:Rs}). Then
\begin{equation}
\underset{\xi \in \mathcal{D}_M}{\sup} \enskip
\frac{1}{M} \left| R_s\left(\xi\right) - R_s^{\circ}\left(\xi\right) \right| \stackrel{\textrm{a.s.}}{\longrightarrow} 0, \quad \textrm{as} \quad M \rightarrow \infty.
\label{eqn:Theorem1}
\end{equation}
$R_s^{\circ}$ denotes the secrecy sum-rate in the large-system regime, given by
\begin{equation}
R_s^{\circ} = K \left[ \log_2 \frac{1+
g\left( \beta,\xi \right)
\frac{\rho + \frac{\rho\xi}{\beta} \left[ 1 + g\left( \beta,\xi \right) \right] ^2 
}{\rho + \left[ 1 + g\left( \beta,\xi \right) \right] ^2}}
{1+\frac{\rho}{ \left( 1+g\left( \beta,\xi \right) \right) ^2}} \right]^+, \quad \textrm{for} \quad \xi \neq 0,
\label{eqn:Rs_large_system}
\end{equation}
with $g \left( \beta,\xi \right) = \frac{1}{2} \left[ \textrm{sgn}(\xi) \cdot \sqrt{ \frac{\left(1-\beta \right)^2}{\xi^2}  +  \frac{2\left(1+\beta\right)}{\xi}  +  1} +  \frac{1-\beta}{\xi}  -  1 \right]$
and
\begin{equation}
  R_s^{\circ}(0) = \lim_{\xi \rightarrow 0} R_s^{\circ}(\xi) = \left\{ 
  \begin{array}{c c l}
    \beta \log_2 \left[ 1+\frac{\left(1-\beta\right)\rho}{\beta} \right] & \quad \textrm{for $\beta \leq 1$} \\
    \beta \left\{ \log_2 \frac{\beta^3\left[\beta+\rho\left(\beta-1\right)\right]}{\left[\beta^2+\rho\left(\beta-1\right)^2\right]^2} \right\}^+ & \quad \text{for $\beta > 1$}\\
  \end{array} \right.
\end{equation}
\end{Theorem}
\begin{IEEEproof}
The proof of Theorem 1 can be found in Appendix A.
\end{IEEEproof}

\subsection{Secrecy Sum-Rate Maximizing Regularization Parameter}

The value of $\xi$ has a significant impact on the large-system secrecy sum-rate $R_s^{\circ}$ in (\ref{eqn:Rs_large_system}). In the following, we derive the regularization parameter $\xi^{\star\circ}$ that maximizes $R_s^{\circ}$.
\begin{Theorem}
Let $\rho>0$, $\beta>0$. Let $\xi^{\star}_M = \underset{\xi \in \mathcal{D}_M}{\arg \max} \enskip R_s(\xi)$ be the optimal regularization parameter in $\mathcal{D}_M$, and denote $R_s^{\star} \triangleq R_s(\xi^{\star}_M)$. Then
\begin{equation}
\frac{1}{M} \left[ R_s^{\star} - R_s^{\star}(\xi^{\star\circ}) \right] \stackrel{\textrm{a.s.}}{\longrightarrow} 0, \quad \textrm{as} \quad M \rightarrow \infty,
\label{eqn:Theorem2}
\end{equation}
where $\xi^{\star\circ} \! \in \! \mathcal{D}_M$ is the optimal large-system regularization parameter, given, for $M$ large enough, by
\begin{equation}
\xi^{\star\circ} = \frac {-2\rho^2\left( 1 - \beta \right)^2 + 6\rho \beta + 2\beta^2 - 2 \left[ \beta \left( \rho+1 \right) -\rho \right] \cdot \sqrt{ \beta^2 \left[ \rho^2 + \rho + 1 \right] - \beta \left[ 2 \rho \left( \rho -1 \right) \right] + \rho^2 } } {6 \rho^2 \left( \beta + 2 \right) + 6 \rho \beta}.
\label{eqn:xi_opt}
\end{equation}
\end{Theorem}
\begin{IEEEproof}
The value of $\xi^{\star\circ}$ can be found by setting the derivative of $R_{s}^{\circ}$ to zero and studying its maxima in each of the intervals which compose the set $\mathcal{D}_M$. Then we have
\begin{align}
0 & \stackrel{a}{\leq} \frac{1}{M} \left[ R_s^{\star} - R_s(\xi^{\star\circ}) \right] = \frac{1}{M} \left[ R_s(\xi^{\star}_M) - R_s^{\circ}(\xi^{\star}_M) + R_s^{\circ}(\xi^{\star}_M) - R_s^{\circ}(\xi^{\star\circ}) + R_s^{\circ}(\xi^{\star\circ}) - R_s(\xi^{\star\circ}) \right] \\
& \stackrel{b}{\leq} \frac{1}{M} \left[ R_s(\xi^{\star}_M) - R_s^{\circ}(\xi^{\star}_M) + R_s^{\circ}(\xi^{\star\circ}) - R_s(\xi^{\star\circ}) \right] \stackrel{c}{\longrightarrow} 0,
\end{align}
where (a), resp. (b), follows from the definition of $\xi^{\star}_M$, resp. $\xi^{\star\circ}$, and (c) follows from Theorem 1. 
\end{IEEEproof}

When $\beta = 1$, the value of $\xi^{\star\circ}$ in (\ref{eqn:xi_opt}) reduces to the one derived in \cite{Geraci12}, given by $\xi^{\star\circ}= \left(3 \rho + 1 + \sqrt{3 \rho + 1}\right)^{-1}$.
We note that the value $\xi^{\star\circ}$ that maximizes the secrecy sum-rate can be negative, and it differs from the value $\xi^{\star\circ}_{\mathrm{ns}} = \beta/\rho$ that maximizes the sum-rate without secrecy requirements \cite{Nguyen09}. Unlike $\xi^{\star\circ}_{\mathrm{ns}}$, which grows unbounded as $\rho \rightarrow 0$, $\xi^{\star\circ}$ is upper bounded by $\xi^{\star\circ}_0 = \lim_{\rho \rightarrow 0} \xi^{\star\circ} = 1 - \frac{\beta}{2}$, 
$\forall \beta >0$, although when $\beta\geq2$ it can be shown that $R_s^{\circ}=0$ irrespective of $\xi$ and $\rho$.
Similarly to $\xi^{\star\circ}_{\mathrm{ns}}$, the value of $\xi^{\star\circ}$ decreases as the SNR increases. In the high-SNR regime, we have $\lim_{\rho \rightarrow \infty} \xi^{\star\circ} - \xi^{\star\circ}_{\infty} = 0$,
where $\xi^{\star\circ}_{\infty}$ approximates the high-SNR behavior of $\xi^{\star\circ}$ and is given by
\begin{equation}
  \xi^{\star\circ}_{\infty} = \left\{ 
  \begin{array}{c c l}
    \frac{\beta}{2\rho} & \quad \textrm{for $\beta < 1$} \\
    \frac{1}{3\rho} & \quad \textrm{for $\beta = 1$} \\
    - \frac{2\left(\beta-1\right)^2}{3\left(\beta+2\right)} + \frac{\beta\left(2-\beta\right)}{2\rho\left(\beta+2\right)}  & \quad \text{for $\beta > 1$}\\
  \end{array} \right.
\label{eqn:xi_opt_large_SNR}
\end{equation}
We then have by the continuous mapping theorem
\begin{equation} \lim_{\rho \rightarrow \infty} \frac{R_s^{\star\circ} - R_s^{\star\circ\infty}}{R_s^{\star\circ}} = 0, \quad \textrm{with} \quad R_s^{\star\circ} \triangleq R_s^{\circ}\left(\xi^{\star\circ}\right) \quad \textrm{and} \quad R_s^{\star\circ\infty} \triangleq \lim_{\rho\rightarrow\infty}R_s^{\circ}(\xi_{\infty}^{\star\circ}).
\label{eqn:Rs_RCI_large_SNR}
\end{equation}

\subsection{Optimal Secrecy Sum-Rate}

By substituting $\xi^{\star\circ}$ from (\ref{eqn:xi_opt}) in (\ref{eqn:Rs_large_system}), it is possible to obtain the optimal secrecy sum-rate $R_s^{\star\circ}$ achievable by RCI precoding in the large-system regime. The secrecy sum-rate $R_s^{\star\circ}$ is a function of $M$, $\beta$ and $\rho$. It was shown in \cite{Geraci12} that for $\beta=1$, $R_s^{\star\circ}$ is always positive and monotonically increasing with the SNR $\rho$. It can be shown that the same is true for $\beta < 1$. However when $\beta > 1$, the secrecy sum-rate does not monotonically increase with $\rho$. It will be shown in Section IV that there is an optimal value of the SNR beyond which the achievable secrecy sum-rate $R_s^{\star\circ}$ starts decreasing, until it becomes zero for large SNR. When $\beta \geq 2$ no positive secrecy sum-rate is achievable at all. 

These results can be explained as follows. In the worst-case scenario, the alliance of cooperating eavesdroppers can cancel the interference, and its received SINR is the ratio between the signal leakage and the thermal noise. In the limit of large SNR, the thermal noise vanishes, and the only means for the transmitter to limit the eavesdropper's SINR is by reducing the signal leakage to zero by inverting the channel matrix. This can only be accomplished when the number of transmit antennas is larger than or equal to the number of users, hence only if $\beta \leq 1$. When $\beta>1$ this is not possible, and no positive secrecy sum-rate can be achieved. When $\beta\geq2$, the eavesdroppers are able to drive the secrecy sum-rate to zero irrespective of $\rho$. This result is expected and consistent with the ones in \cite{Khisti10I}.

\newtheorem{Remark}{Remark}
\begin{Remark}
In order for Theorem 1 to hold with $\xi=\xi^{\star\circ}$, it is sufficient that $\xi^{\star\circ} \in \mathcal{D}_M$. Since $\xi^{\star\circ}$ in (\ref{eqn:xi_opt}) depends on $\beta$ and $\rho$, so does the accuracy of $R_s^{\star\circ}$ for finite $M$. We can distinguish the two following cases. (i) When $\beta\neq1$, we have $\xi^{\star\circ} \in \mathcal{D}_M$ $\forall \rho$, and the approximation is accurate uniformly on $\rho$. (ii) When $\beta = 1$, $\xi^{\star\circ} \in \mathcal{D}_M$ for all finite $\rho$; if $\rho \rightarrow \infty$, then it is required that $M=\mathcal{O}(\rho^{2+\epsilon})$, for some $\epsilon>0$, otherwise the approximation gets weaker as $\rho \rightarrow \infty$ for $M$ fixed. This means for instance that the approximation with $M=10$ and $\rho=17$dB is as accurate as the approximation with $M=40$ and $\rho=20$dB.
\end{Remark}

\subsection{Optimal Network Load}

\begin{figure}
\centering
\includegraphics[width=\columnwidth]{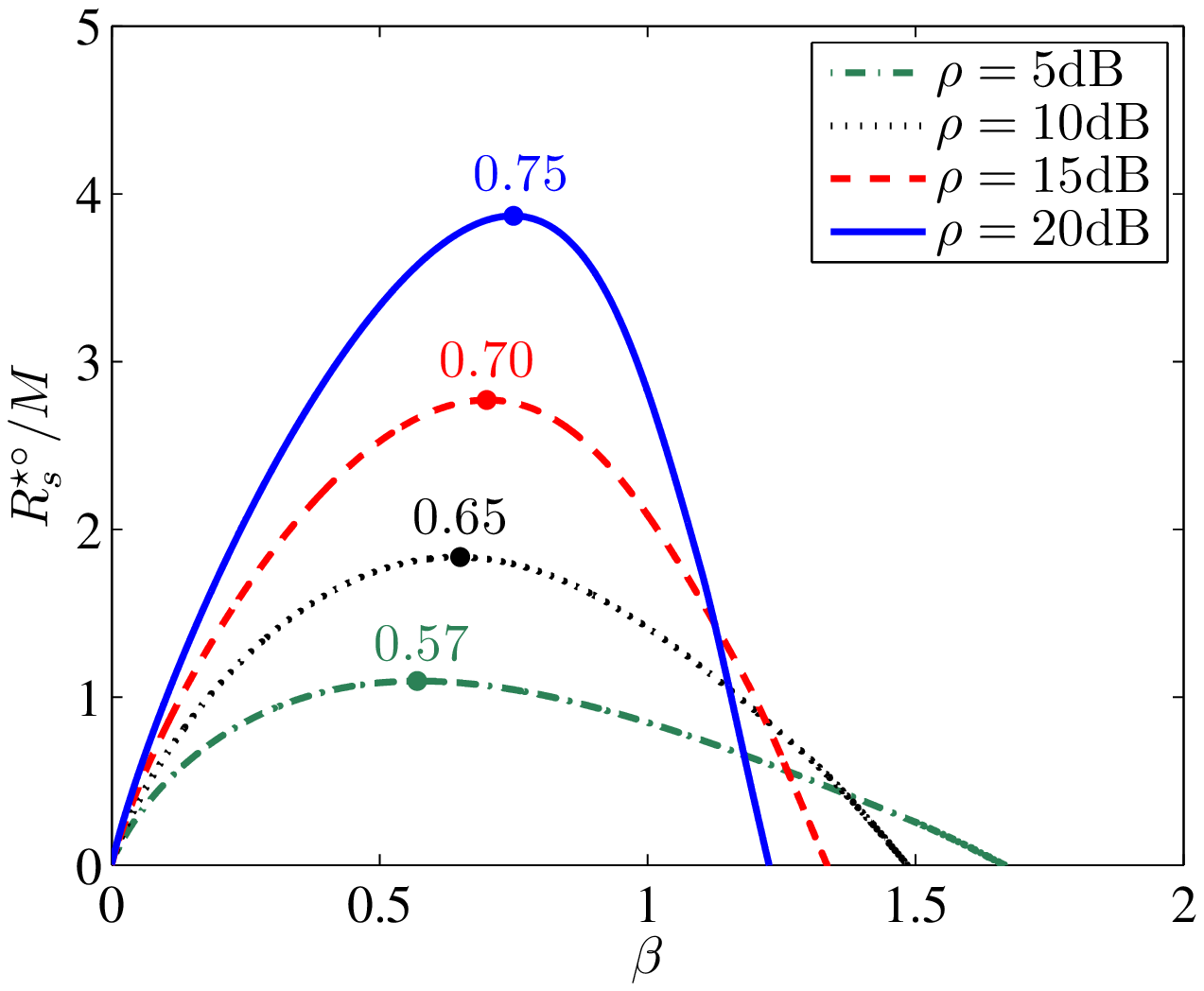}
\caption{Asymptotic secrecy sum-rate per transmit antenna as a function of $\beta$ for RCI precoding. The value of $\beta^{\star\circ}$ is indicated next to each curve.}
\label{fig:Rs_vs_beta}
\end{figure}

Fig. \ref{fig:Rs_vs_beta} depicts the per-antenna secrecy sum-rate $R_s^{\star\circ}/M$ as a function of the network load $\beta$, for several values of the SNR. We denote by $\beta^{\star\circ}$ the value of $\beta \in \mathbb{R}^+$ that maximizes the per-antenna secrecy sum-rate.
It is possible to see from Fig. \ref{fig:Rs_vs_beta} that the value of $\beta^{\star\circ}$ falls between $0$ and $1$, and that it is an increasing function of the SNR. A closed-form approximation for $\beta^{\star\circ}$ in the large-SNR regime is given in the following.
\newtheorem{Proposition}{Proposition}
\begin{Proposition}
In the limit of large SNR, the value $\tilde{\beta}^{\star\circ}$ of the optimal network load can be found by solving the following fixed point equation
\begin{equation}
\tilde{\beta}^{\star\circ} = \rho \left( 1-\tilde{\beta}^{\star\circ} \right) e^{-\frac{1}{1-\tilde{\beta}^{\star\circ}}},
\label{eqn:beta_opt}
\end{equation}
and the network load $\tilde{\beta}^{\star\circ}$ tends to one for large SNR.
\end{Proposition}
\begin{IEEEproof}
From (\ref{eqn:Rs_RCI_large_SNR}), we have that $R_s^{\star\circ\infty}$ approximates $R_s^{\star\circ}$ in the large-SNR regime. We then obtain (\ref{eqn:beta_opt}) by noticing that it must be $\tilde{\beta}^{\star\circ} \in [0,1]$, and by setting $\partial (R_s^{\star\circ\infty}/M)/\partial\beta=0$.
\end{IEEEproof}

\subsection{Numerical Results}

\begin{figure}
\centering
\includegraphics[width=\columnwidth]{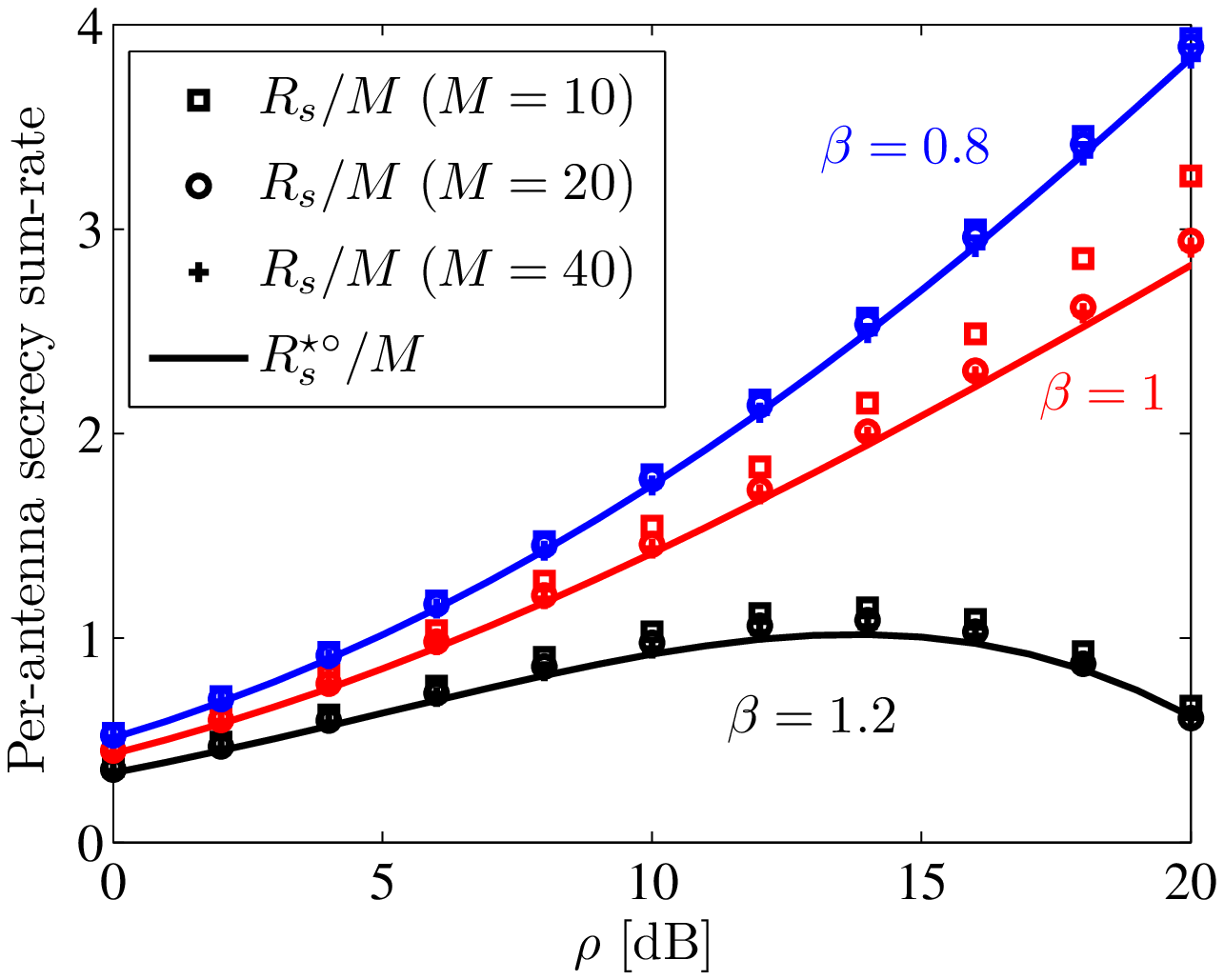}
\caption{Comparison between the secrecy sum-rate with RCI precoding in the large-system regime (\ref{eqn:Rs_large_system}) and the simulated ergodic secrecy sum-rate for finite $M$. Three sets of curves are shown, each one corresponds to a different value of $\beta$.}
\label{fig:Rs_analysis_vs_sim}
\end{figure}

Fig. \ref{fig:Rs_analysis_vs_sim} compares the secrecy sum-rate $R_s^{\star\circ}$ of the RCI precoder from the large-system analysis to the simulated ergodic secrecy sum-rate $R_s$ with a finite number of users, for different values of $\beta$. The value of $R_s^{\star\circ}$ was obtained by (\ref{eqn:Rs_large_system}) with $\xi^{\star\circ}$ as in (\ref{eqn:xi_opt}). The value of $R_s$ was obtained by using the regularization parameter that maximizes the average secrecy sum-rate. We observe that when $\beta = 0.8$ and when $\beta=1.2$ the large-system analysis is accurate for all values of $M$ and SNR. When $\beta = 1$, the analysis is accurate at low SNR for all values of $M$, and for high SNR larger values of $M$ are required to increase the accuracy. 
The previous observations are consistent with Remark 1.

\begin{figure}
\centering
\includegraphics[width=\columnwidth]{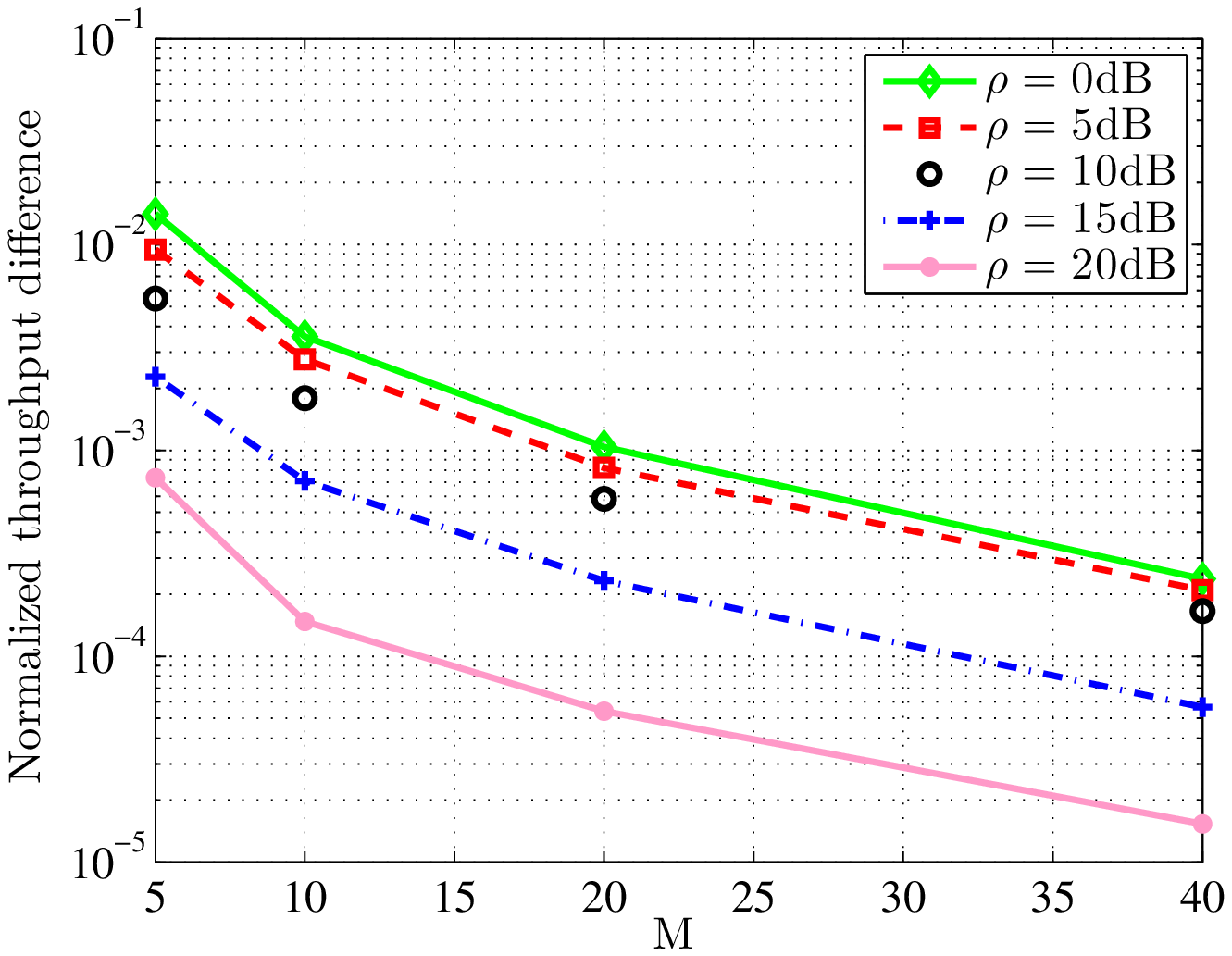}
\caption{Mean normalized secrecy sum-rate difference between $R_s^{\star}$ (obtained using the optimal $\xi^{\star}_M$) and $R_s(\xi^{\star\circ})$ (obtained with $\xi^{\star\circ}$ from large-system analysis), for $\beta=0.8$ and various values of the SNR.}
\label{fig:Rs_loss_vs_M}
\end{figure}

Fig. \ref{fig:Rs_loss_vs_M} shows that using the regularization parameter $\xi^{\star\circ}$, obtained from large-system analysis, does not
cause a significant loss in the secrecy sum-rate compared
to using $\xi^{\star}_M$, optimized for
each channel realization. The figure shows the normalized secrecy sum-rate difference $\left(R_s^{\star}-R_s(\xi^{\star\circ})\right)/R_s^{\star}$, simulated for finite-size systems, $\beta=0.8$ and various values of the SNR.
The value of $R_s^{\star}$ was obtained by using $\xi^{\star}_M$, whereas $R_s(\xi^{\star\circ})$ was obtained by using $\xi^{\star\circ}$. We observe that the average normalized secrecy sum-rate
difference is less than $2\%$ for all values of $M$ and $\rho$. As a result, one can avoid the calculation of $\xi^{\star}_M$ for every channel realization, and $\xi^{\star\circ}$ can be used with
only a small loss of performance.

\begin{figure}
\centering
\includegraphics[width=\columnwidth]{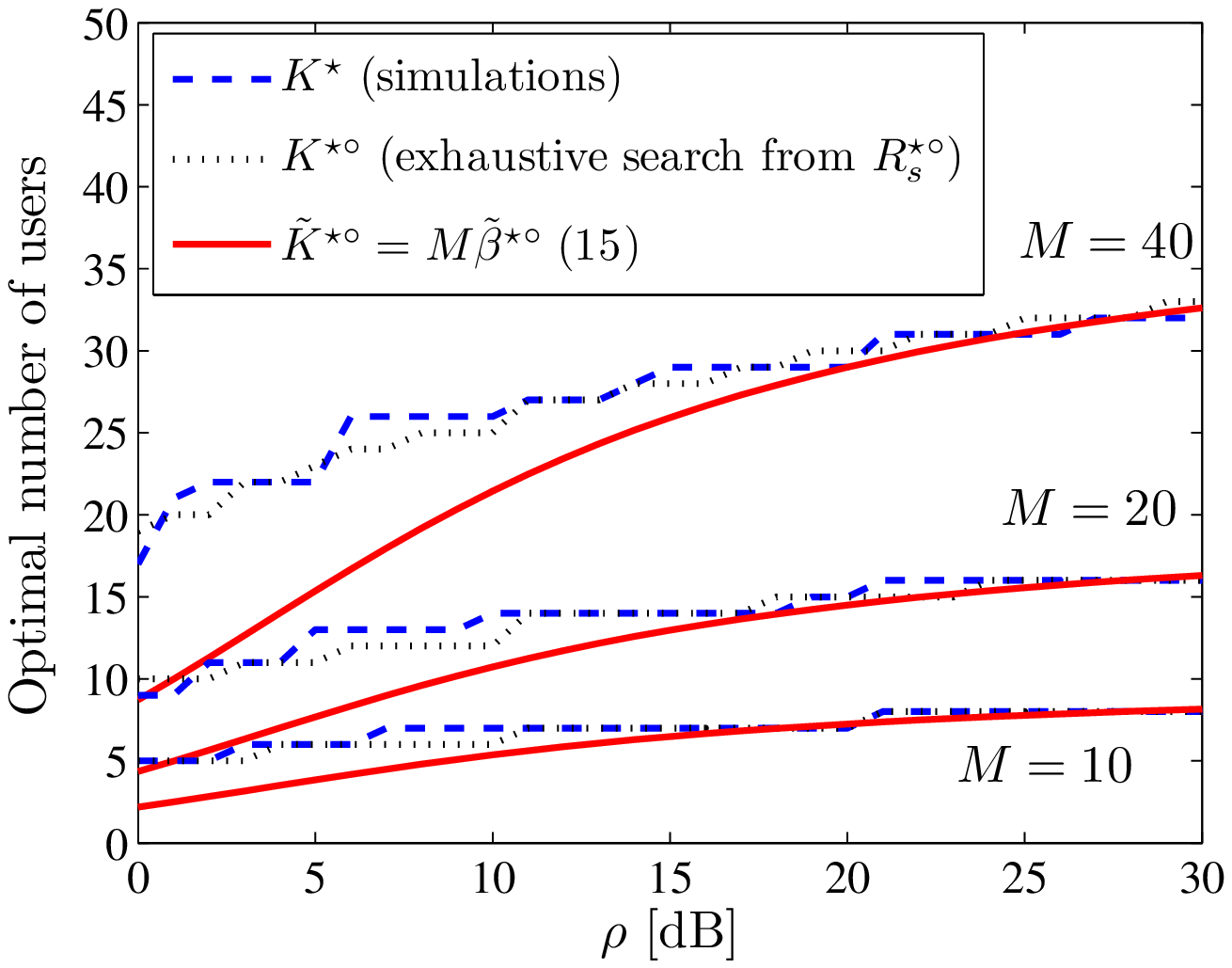}
\caption{Comparison between $K^{\star}$ (obtained via simulations), $K^{\star\circ}$ (obtained via exhaustive search and large-system analysis), and the closed-form approximation $\tilde{K}^{\star\circ}$ (obtained via large-system and large-SNR analysis).}
\label{fig:K_opt}
\end{figure}

Fig. \ref{fig:K_opt} shows the optimal number of users $K^{\star}$ obtained via simulations, for $M = 10$, $20$, and $40$ antennas. This is compared to $K^{\star\circ}$, obtained from an exhaustive search on the the large-system rate $R_s^{\star\circ}$, and to the closed-form approximation $\tilde{K}^{\star\circ}$, obtained from (\ref{eqn:beta_opt}) in the high-SNR regime. We note that $K^{\star\circ}$ is accurate across the whole range of SNR, whereas $\tilde{K}^{\star\circ}$ is accurate for medium-to-large values of the SNR.
\section{RCI Precoder with Power Reduction}

We have found that for $\beta >1$ the RCI precoder performs poorly in the high-SNR regime. In this section, we first derive the optimal value of the SNR $\rho^{\star\circ}$ that maximizes the achievable secrecy sum-rate $R_s^{\star\circ}$ for $\beta>1$. We then propose a linear precoder based on RCI and power reduction which significantly increases the high-SNR secrecy sum-rate for $1<\beta<2$.

\subsection{Optimal Transmit SNR}

When $1 < \beta < 2$, there is an optimal value of the transmit SNR $\rho^{\star\circ}$, provided in the following.
\begin{Proposition}
The value of the SNR $\rho^{\star\circ}$ that maximizes the secrecy sum-rate $R_s^{\star\circ}$ for $1 < \beta < 2$, and the corresponding maximum value  of $R_s^{\star\circ}$ are respectively given by
\begin{equation}
\rho^{\star\circ} = \underset{\rho}{\arg \max} \enskip R_s^{\star\circ} \left(\rho\right) = \frac{\beta \left( 2 - \beta \right)}{\left( \beta - 1 \right) ^2} \quad \textrm{and} \quad R_s^{\star\circ} \left(\rho^{\star\circ}\right) = K \log_2 \frac{\beta^2}{4 \left( \beta - 1 \right)}.
\label{eqn:rho_opt}
\end{equation}
\label{prop:rho_opt}
\end{Proposition}
\begin{IEEEproof}
If $1 < \beta < 2$, then $\rho^{\star\circ}$ is the only stationary point of $R_{s}^{\star\circ}$, which can be found by setting its derivative $\partial R_{s}^{\star\circ} / \partial \rho$ to zero. We note that $\lim_{\rho \rightarrow \rho^{\star\circ}} \xi^{\star\circ} = 0$. Therefore, $R_s^{\star\circ} \left(\rho^{\star\circ}\right)$ can be obtained by considering $\rho \rightarrow \rho^{\star\circ}$ and $\xi \rightarrow 0$ in (\ref{eqn:Rs_large_system}) and after several algebraic manipulations.
\end{IEEEproof}

\subsection{Power Reduction Strategy}

We now propose a power reduction strategy to prevent the secrecy sum-rate from decreasing at high SNR, for $1<\beta<2$. This is achieved by reducing the transmit power, and therefore reducing the SNR to the value $\rho^{\star\circ}$ that maximizes the secrecy sum-rate. We denote this scheme by the RCI precoder with power reduction (RCI-PR), whose precoding matrix $\mathbf{W}_{\textrm{PR}}$ is given by
\begin{equation}
\mathbf{W}_{\textrm{PR}} = \left\{ 
  \begin{array}{l l l}
    \frac{1}{\sqrt{\gamma}} \mathbf{H}^H \left( \mathbf{H H}^H + M \xi \mathbf{I}_K \right) ^{-1} & \quad \textrm{for $\beta \leq 1$} \\
    \frac{1}{\sqrt{r \gamma}} \left( \mathbf{H}^H \mathbf{H} + M \xi \mathbf{I}_M \right) ^{-1} \mathbf{H}^H & \quad \textrm{for $1 < \beta < 2$} \\
    0 & \quad \text{for $\beta \geq 2$}\\
  \end{array} \right.
\label{eqn:RCI_PR}
\end{equation}
where $r=\max \left( \frac{\rho}{\rho^{\star\circ}}, 1 \right)$ is the power reduction constant used for $1 < \beta < 2$,
and where $\xi$ is chosen from (\ref{eqn:xi_opt}) evaluated with an SNR of $\min ( \rho, \rho^{\star\circ})$. We note that (\ref{eqn:RCI_PR}) generalizes the RCI precoder in (\ref{eqn:RCI_precoder}) to the case when the power reduction strategy is employed.

\begin{Remark}
We note from (\ref{eqn:xi_opt}) that $\xi^{\star\circ}(\rho^{\star\circ})=0$. Therefore when $\rho \geq \rho^{\star\circ}$, the optimal value of $\xi$ for the RCI-PR precoder is zero, and it reduces to a CI-PR precoder. Even if $\beta > 1$, it is still possible to calculate $\mathbf{W}_{\textrm{PR}}$ by expressing it as in (\ref{eqn:RCI_PR}) for $1 < \beta < 2$.
\end{Remark}

We denote by $R_{s,\textrm{PR}}^{\circ}$ the large-system secrecy sum-rate achievable by the proposed RCI-PR precoder (\ref{eqn:RCI_PR}). 
The following theorem provides a high-SNR approximation of $R_{s,\textrm{PR}}^{\circ}$.

\begin{Theorem}
In the high-SNR regime, we have $\lim_{\rho \rightarrow \infty} \frac{R_{s,\textrm{PR}}^{\circ} - R_{s,\textrm{PR}}^{\circ\infty}}{R_{s,\textrm{PR}}^{\circ}} = 0$,
where $R_{s,\textrm{PR}}^{\circ\infty}$ approximates the large-system secrecy sum-rate $R_{s,\textrm{PR}}^{\circ}$ achieved by the RCI-PR precoder, and it is given by
\begin{equation}
  R_{s,\textrm{PR}}^{\circ\infty} = \left\{ 
  \begin{array}{l c l}
     K \log_2 {\frac{1-\beta}{\beta}} + K \log_2 {\rho} & \quad \textrm{for $\beta < 1$} \\
    \frac{K}{2} \log_2 \frac{27}{64} + \frac{K}{2} \log_2 {\rho} & \quad \text{for $\beta = 1$} \\
    K \log_2 \frac{\beta^2}{4 \left( \beta - 1 \right)} & \quad \text{for $1 < \beta < 2$}\\
    0 & \quad \text{for $\beta \geq 2$}\\
  \end{array} \right.
\label{eqn:Rs_PR_large_SNR}
\end{equation}
\end{Theorem}
\begin{IEEEproof}
When $\beta\leq1$, the RCI-PR precoder reduces to the optimal RCI precoder. Therefore, in this case we have $R_{s,\textrm{PR}}^{\circ\infty}=R_s^{\star\circ\infty}$, with the latter defined in (\ref{eqn:Rs_RCI_large_SNR}). The value of (\ref{eqn:Rs_PR_large_SNR}) for $1 < \beta < 2$ is obtained by noting that for large SNR, RCI-PR forces $\rho = \rho^{\star\circ}$, and by using Proposition \ref{prop:rho_opt}. The value for $\beta \geq 2$ arises from the fact that no positive secrecy sum-rate is achievable in such a condition, therefore the RCI-PR precoder (\ref{eqn:RCI_PR}) transmits zero power.
\end{IEEEproof}

From (\ref{eqn:Rs_PR_large_SNR}) we can conclude that the behavior of our proposed RCI-PR precoder can be classified into four regions. When $\beta < 1$, any secrecy sum-rate can be achieved, as long as the transmitter has enough power available, and the secrecy sum-rate scales linearly with the factor $K$. When $\beta = 1$, the linear scaling factor reduces to $K / 2$. When $1 < \beta < 2$, the cooperating eavesdroppers have more antennas than the transmitter, and thus they can limit the achievable secrecy sum-rate regardless of how much power is available at the transmitter. When $\beta \geq 2$, the eavesdroppers are able to prevent secret communications, and the secrecy sum-rate is zero even if unlimited power is available.

\subsection{Numerical Results}

\begin{figure}
\centering
\includegraphics[width=\columnwidth]{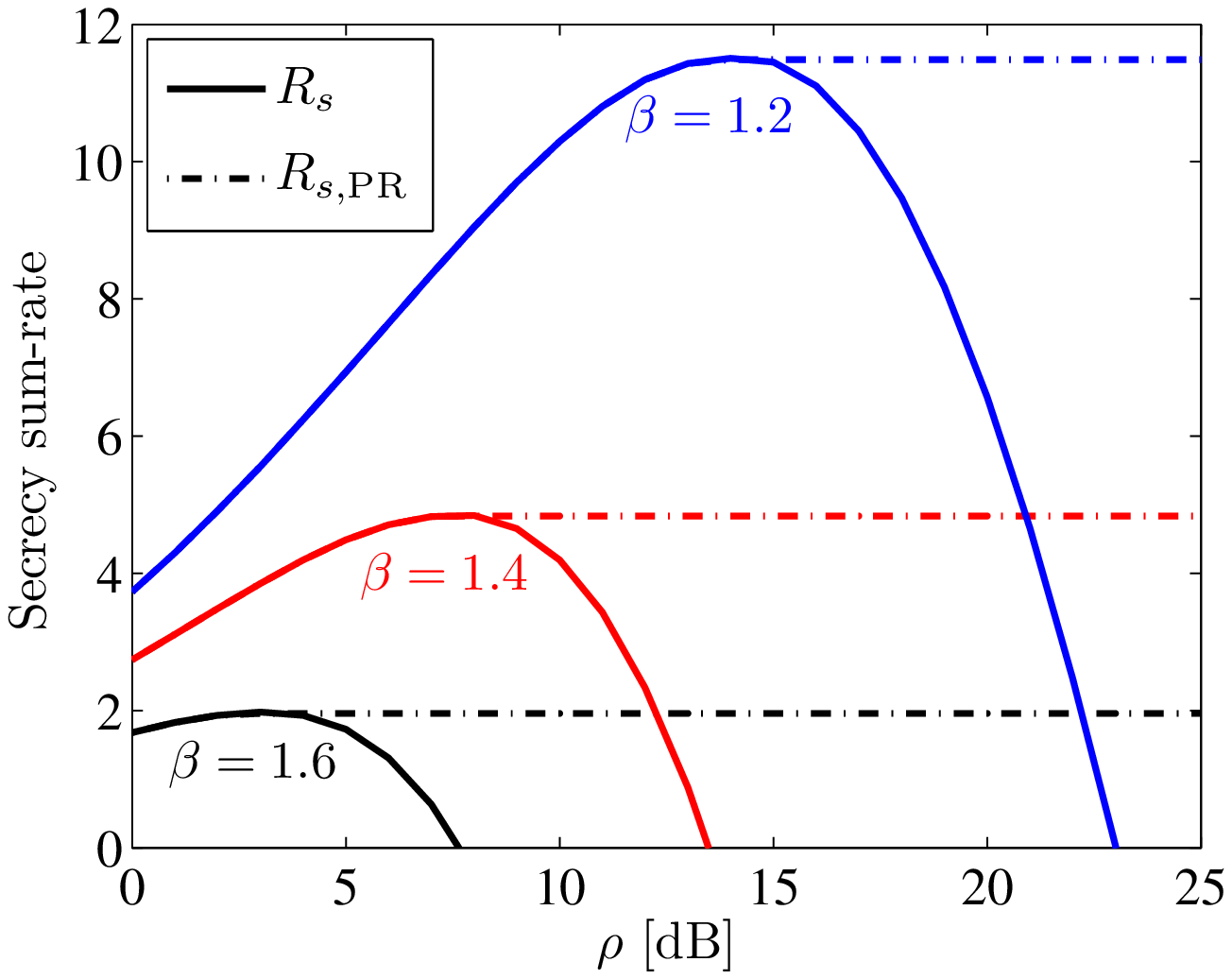}
\caption{Comparison between the ergodic secrecy sum-rates $R_s$ and $R_{s,\textrm{PR}}$ achieved by the RCI precoder and by the proposed RCI-PR precoder, respectively, for $M=10$ transmit antennas. Three values of $\beta$ are considered: $1.2$, $1.4$ and $1.6$, corresponding to $K = 12$, $14$ and $16$ users.}
\label{fig:power_reduction}
\end{figure}

Fig. \ref{fig:power_reduction} shows the simulated ergodic secrecy sum-rates with and without the power reduction strategy for $M=10$ transmit antennas and three values of $\beta > 1$. The figure shows that the proposed RCI-PR precoder in (\ref{eqn:RCI_PR}) increases the secrecy sum-rate compared to the RCI precoder in (\ref{eqn:RCI_precoder}). By using the proposed power reduction strategy, it is possible to prevent the secrecy sum-rate from decreasing at large values of the SNR $\rho$. For large $\rho$, the achieved secrecy sum-rate equals the maximum across all values of $\rho$. Moreover, this is achieved by using a lower transmit power, and the amount of power saved equals $10 \log_{10} r^{-1}$ dB.
\section{Performance Comparison}

In this section, we first compare the secrecy sum-rate $R_{s,\textrm{PR}}^{\star\circ}$ achieved by the proposed RCI-PR precoder to the sum-rate $R^{\star\circ}$ achieved by the optimized RCI precoder without secrecy requirements, in the large-system regime. The gap between $R_{s,\textrm{PR}}^{\star\circ}$ and $R^{\star\circ}$ represents the \emph{secrecy loss}, i.e. how much the secrecy requirements cost in terms of the achievable sum-rate. Furthermore, we compare the per-user secrecy rate achieved by the proposed precoder to the secrecy capacity $C_{s,\textrm{SU}}$ of a single-user MISOME wiretap channel \cite{Khisti10I}. The gap between $R_{s,\textrm{PR}}^{\star\circ} / K$ and $C_{s,\textrm{SU}}$ represents a \emph{multi-user loss}, i.e. the loss due to the requirement of serving multiple users at the same time.

\subsection{Secrecy Loss}

The secrecy sum-rate $R_{s,\textrm{PR}}^{\star\circ}$ is obtained by using the precoder in (\ref{eqn:RCI_PR}). The optimal sum-rate $R^{\star\circ}$ without secrecy requirements is obtained by using the precoder in (\ref{eqn:RCI_precoder}), and it is given by \cite{Nguyen09}
\begin{equation}
R^{\star\circ} = K \log_2 { \left[ 1 + g \left( \beta, \xi^{\star\circ}_{\textrm{ns}} \right) \right] },
\end{equation}
with $\xi^{\star\circ}_{\textrm{ns}} = \beta/\rho$.
Similarly to the secrecy sum-rate, there is an optimal value for the ratio $\beta$ that maximizes the per-antenna sum-rate $R^{\star\circ} / M$ without secrecy requirements \cite{Nguyen09,HochwaldAllerton02,Wagner12}.
It is easy to show that $R^{\star\circ} \geq 0$ for all values of $\beta$ and $\rho$, with equality only for $\rho = 0$, and that $R^{\star\circ}$ tends to zero as $\beta \rightarrow \infty$. Hence, there is no limit to the number of users per transmit antenna $\beta$ that the system can accommodate with a non-zero sum-rate. However if we impose the secrecy requirements, the secrecy sum-rate $R_{s,\textrm{PR}}^{\star\circ}$ is zero for $\beta \geq 2$. Therefore, introducing the secrecy requirements will limit the number of users that can be served with a non-zero rate to two times the number of transmit antennas.

We now compare the secrecy sum-rate $R_{s,\textrm{PR}}^{\star\circ}$ to the sum-rate $R^{\star\circ}$ in the limit of large SNR. Again by using the regularization parameter $\xi^{\star\circ}_{\textrm{ns}} = \beta/\rho$ we obtain $\lim_{\rho \rightarrow \infty} \frac{R^{\star\circ} - R^{\star\circ\infty}}{R^{\star\circ}} = 0$,
with
\begin{equation}
  R^{\star\circ\infty} = \left\{ 
  \begin{array}{l c l}
    K \log_2 {\frac{1-\beta}{\beta}} + K \log_2 {\rho} &  \quad \textrm{for $\beta < 1$} \\
    \frac{K}{2} \log_2 {\rho} &  \quad \text{for $\beta = 1$}\\
    K \log_2 {\frac{\beta}{\beta-1}} &  \quad \text{for $\beta > 1$}\\
  \end{array} \right.
\label{eqn:R_RCI_large_SNR}
\end{equation}
By comparing (\ref{eqn:R_RCI_large_SNR}) to (\ref{eqn:Rs_PR_large_SNR}), we can draw the following conclusions regarding the large-SNR regime. If the number of transmit antennas $M$ is larger than the number of users $K$, then $R_{s,\textrm{PR}}^{\star\circ\infty} = R^{\star\circ\infty}$ and the secrecy requirements do not decrease the sum-rate of the network. Therefore by using the RCI-PR precoder in (\ref{eqn:RCI_PR}), one can achieve secrecy while maintaining the same sum-rate, i.e. there is no secrecy loss. If $M = K$, then the secrecy loss is $\frac{1}{2}\log_2(\frac{64}{27})\approx0.62$ bits per user, but the linear scaling factor $K/2$ remains unchanged. Alternatively, one can achieve secrecy while maintaining the same sum-rate, by increasing the transmit power by a factor $64/27 \approx 3.75$dB. If $M < K < 2M$, then the secrecy loss is ($2-\log_2 \beta$) bits per user, but the proposed precoder transmits a lower power, which is always upper bounded by $\frac{\beta\left(2-\beta\right)}{\left(\beta-1\right)^2}$. Finally if $K \geq 2M$, then the secrecy requirements force the sum-rate to zero, whereas the sum-rate $R^{\star\circ}$ remains positive, though it also tends to zero for large $\beta$. We finally note that, when there is no secrecy constraint, user scheduling can be used to achieve a higher multiplexing gain. This is not possible in the BCC, since discarding users does not prevent them from eavesdropping.

\subsection{Multi-User Loss}

We now consider the \emph{multi-user loss}, i.e. the loss due to the interference caused by the presence of multiple users in the system. This is given by the gap between the per-user secrecy rate $R_{s,\textrm{PR}}^{\star\circ}/K$ achieved by the proposed RCI-PR precoder and the secrecy capacity $C_{s,\textrm{SU}}$ of the MISOME wiretap channel, where one user is served at a time and the remaining users can eavesdrop \cite{Khisti10I}. We compare $R_{s,\textrm{PR}}^{\star\circ}/K$ to $C_{s,\textrm{SU}}$ in the large-SNR regime. The former is obtained from (\ref{eqn:Rs_PR_large_SNR}).
The value of $C_{s,\textrm{SU}}$ was obtained in \cite{Khisti10I}, and for large SNR we have $\lim_{\rho \rightarrow \infty} \frac{C_{s,\textrm{SU}} - C_{s,\textrm{SU}}^{\infty}}{C_{s,\textrm{SU}}} = 0$,
where
\begin{equation}
  C_{s,\textrm{SU}}^{\infty} = \left\{ 
  \begin{array}{l c l}
    \log_2 {\rho} & \quad \textrm{for $\beta < 1$} \\
    \frac{1}{2} \log_2 \rho & \quad \text{for $\beta = 1$} \\
    \log_2 \frac{1}{\left( \beta - 1 \right)} & \quad \text{for $1 < \beta < 2$}\\
    0 & \quad \text{for $\beta \geq 2$}\\
  \end{array} \right.
\label{eqn:Cs_Khisti}
\end{equation}

We note that in $C_{s,\textrm{SU}}$ from \cite{Khisti10I} a single-user system is considered. Therefore, only one message is transmitted to one legitimate user, and the user does not experience any interference. By comparing (\ref{eqn:Cs_Khisti}) to $R_{s,\textrm{PR}}^{\star\circ}/K$, we can conclude that the multi-user loss is $\log_2\frac{1-\beta}{\beta}$ and $0.62$ bits per user for $\beta<1$ and $\beta=1$, respectively. Hence for $\beta \leq 1$, the proposed RCI-PR precoder achieves a per-user secrecy rate which has the same linear scaling factor as the secrecy capacity of a single-user system with no interference. When $1 < \beta < 2$, the proposed precoder suffers a multi-user loss of $( 2 - 2 \log_2 \beta)$ bits, but again it has the advantage of transmitting a limited power.

\subsection{Numerical Results}

\begin{figure}
\centering
\includegraphics[width=\columnwidth]{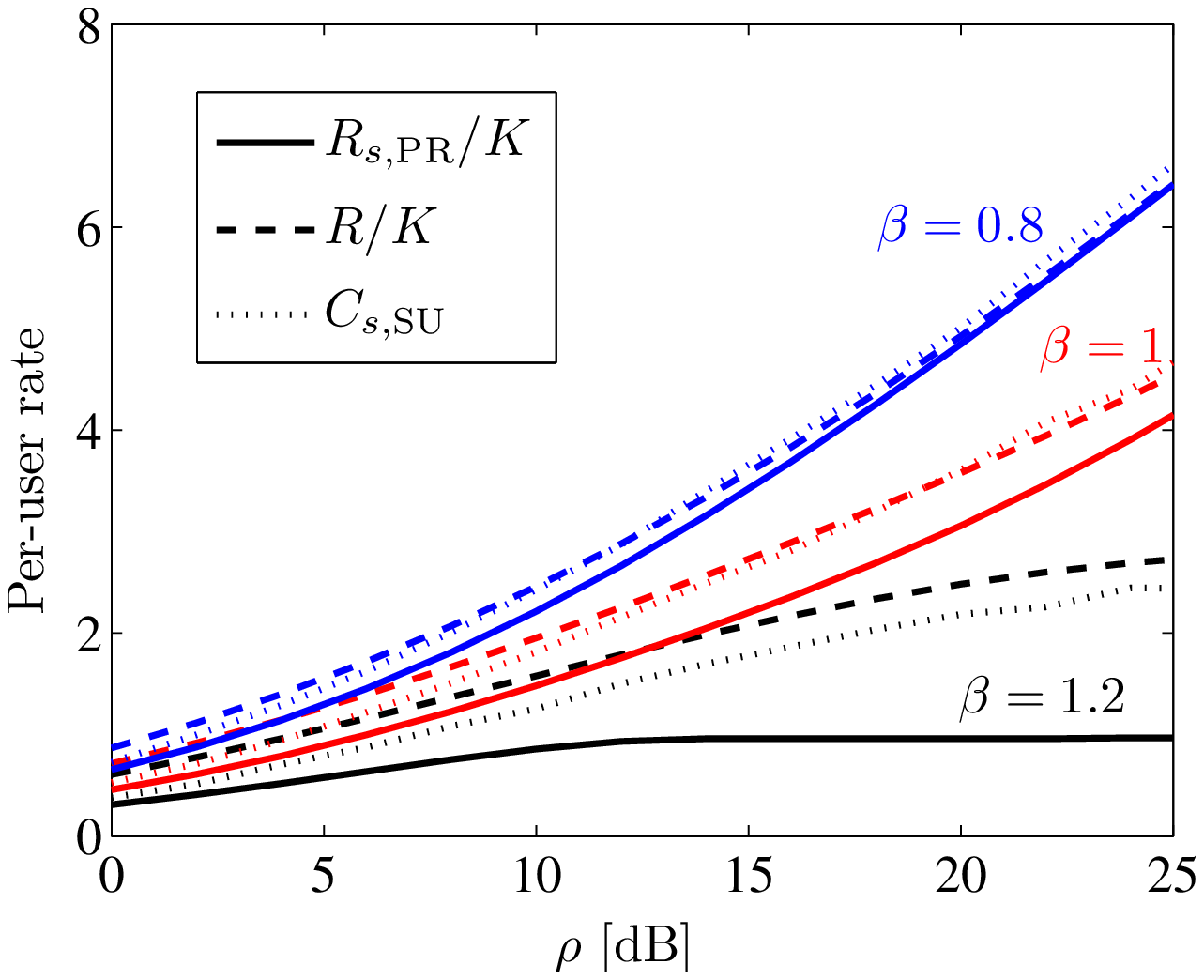}
\caption{Comparison between the simulated ergodic per-user secrecy rate with RCI-PR (solid) and the two upper bounds: (i) per-user rate without secrecy requirements (dashed) and (ii) MISOME secrecy capacity (dotted), for $K=12$ users. Three values of $\beta$ are considered: $0.8$, $1$, and $1.2$, corresponding to $M=15$, $12$ and $10$ antennas.}
\label{fig:Rs_vs_R_MISOME_sim}
\end{figure}

In Fig. \ref{fig:Rs_vs_R_MISOME_sim} we compare the simulated per-user ergodic sum-rate $R_{s,\textrm{PR}}/K$ of the RCI-PR precoder to the sum-rate $R/K$ of the RCI precoder without secrecy requirements. These were obtained by using the regularization parameters $\xi^{\star\circ}$ and $\xi_{\textrm{ns}}^{\star\circ}$, respectively.
For $\beta<1$, the difference between $R_{s,\textrm{PR}}/K$ and $R/K$ becomes negligible at large SNR, and secrecy can be achieved without additional costs. For $\beta=1$, the two curves tend to have the same slope at large SNR, but there is a residual gap between them. Therefore, secrecy can be achieved at a lower sum-rate. We note that in order to achieve secrecy without decreasing the sum-rate, the required additional power is less than $4$dB at all SNRs. For $1 < \beta < 2$, the sum-rate $R$ tends to saturate for large SNR, and so does the secrecy sum-rate $R_{s,\textrm{PR}}$. In the simulations, for $\beta=1.2$ and $\rho=25$dB, the gap is about $1.79$ bits, close to $2-\log_2 \beta \approx 1.74$ bits. Moreover, we note that the proposed precoder saves $92\%$ of the transmit power. The gap is smaller for smaller values of the SNR, e.g. it reduces to about $0.72$ bits when we set the transmit power to $10$dB.

Fig. \ref{fig:Rs_vs_R_MISOME_sim} also shows the simulated secrecy capacity $C_{s,\textrm{SU}}$ of the MISOME wiretap channel. For $\beta\leq1$, the RCI-PR precoder achieves a per-user secrecy rate which has the same linear scaling factor as $C_{s,\textrm{SU}}$. When $1 < \beta < 2$, also $C_{s,\textrm{SU}}$ saturates at high SNR. In particular, for $\beta=1.2$ and $\rho=25$dB, the gap with the RCI-PR precoder is about $1.47 \approx 2-2\log_2 \beta$ bits, but the RCI-PR precoder saves $92\%$ of the power. The gap is smaller for smaller values of the SNR, e.g. it reduces to about $0.4$ bits when we set the transmit power to $10$dB. All these numerical results confirm the ones obtained from the large-system analysis.

\section{Imperfect Channel State Information}

In the previous sections, we proposed a linear precoder for the case when perfect CSI is available at the transmitter. In this section, we consider a more realistic scenario where only an estimation of the channel is available at the transmitter, and we obtain a deterministic equivalent for the secrecy sum-rate achievable by RCI precoding. We then analyze the performance of our proposed RCI-PR precoder, and we determine how the CSI estimation error must scale with the SNR and how many feedback bits are required to maintain a given rate gap to the case with perfect CSI.

\subsection{Secrecy Sum-Rates in the Presence of Channel Estimation Error}

We model the relation between the channel $\mathbf{H}$ and the estimated channel $\mathbf{\hat{H}}$ as $\mathbf{H} = \mathbf{\hat{H}} + \mathbf{E}$,
where the matrix $\mathbf{E}$ represents the channel estimation error, and it is independent from $\mathbf{\hat{H}}$. The entries of $\mathbf{\hat{H}}$ and $\mathbf{E}$ are i.i.d. complex Gaussian random variables with zero mean and variances $1-\tau^2$ and $\tau^2$, respectively. The value of $\tau \in [0,1]$ depends on the quality and technique used for channel estimation, and it is the same for all users. When $\tau=0$ the CSI is perfectly known, whereas $\tau=1$ corresponds to the case when no CSI is available at all.

The transmitter uses the knowledge of $\mathbf{\hat{H}}$ to obtain the RCI precoding matrix $\mathbf{\hat{W}}$, given by
\begin{equation}
\mathbf{\hat{W}} = \frac{1}{\sqrt{\hat{\gamma}}} \mathbf{\hat{H}}^H \left( \mathbf{\hat{H}} \mathbf{\hat{H}}^H + M \xi \mathbf{I} \right) ^{-1} = \frac{1}{\sqrt{\hat{\gamma}}} \left( \mathbf{\hat{H}}^H \mathbf{\hat{H}} + M \xi \mathbf{I} \right) ^{-1} \mathbf{\hat{H}}^H
\end{equation}
where $\hat{\gamma} = \textrm{tr} \left\{ \left( \mathbf{\hat{H}}^H \mathbf{\hat{H}} + M \xi \mathbf{I} \right) ^{-2} \mathbf{\hat{H}}^H \mathbf{\hat{H}} \right\}$ is the power normalization costant in the presence of CSI error. 
A secrecy sum-rate achievable by RCI precoding in the presence of channel estimation error with variance $\tau^2$ is given by
\begin{equation}
\hat{R}_s = \sum_{k=1}^{K} \left[ \log_2 \Big( 1 + \hat{\mathrm{SINR}}_{k} \Big) - \log_2 \Big( 1 + \hat{\mathrm{SINR}}_{\widetilde{k}} \Big) \right]^+,
\label{eqn:Rs_imperfect_CSI_finite}
\end{equation}
with
\begin{equation}
\hat{\mathrm{SINR}}_{k} = \frac {\rho \left| \mathbf{h}_k^H \hat{\mathbf{w}}_k \right| ^2} {1 + \rho \sum_{j \neq k} {\left| \mathbf{h}_k^H \hat{\mathbf{w}}_j \right| ^2} } \quad \textrm{and} \quad \hat{\mathrm{SINR}}_{\widetilde{k}} = \rho \left\| \mathbf{H}_k \hat{\mathbf{w}}_{k} \right\| ^2.
\label{eqn:SINR_hat}
\end{equation}
The deterministic equivalent of $\hat{R_s}$ in the presence of channel estimation error is given in the following.
\begin{Theorem}
Let $\rho>0$, $\beta>0$, and $\xi \in \mathcal{D}_M$. Let $\hat{R}_s$ be the secrecy sum-rate in the presence of channel estimation error with variance $\tau^2$, defined in (\ref{eqn:Rs_imperfect_CSI_finite}). Define $\tilde{\rho} \triangleq \frac{\rho\left(1-\tau^2\right)}{\rho \tau^2 + 1}$ and $\tilde{\xi} \triangleq \frac{\xi}{1 - \tau^2}$. Then $\frac{1}{M} \left( \hat{R}_s - \hat{R}_s^{\circ} \right) \stackrel{M \rightarrow \infty}{\longrightarrow} 0$
almost surely, where $\hat{R}_s^{\circ}$ is the large-system secrecy sum-rate in the presence of CSI error, given by
\begin{equation}
\hat{R}_s^{\circ} = K \left[ \log_2 \frac{1+ g ( \beta,\tilde{\xi} ) \frac{\tilde{\rho} + \frac{ \tilde{\xi} \tilde{\rho}}{\beta} \left[ 1 + g ( \beta,\tilde{\xi} ) \right]^2}{\tilde{\rho} + \left[ 1 + g ( \beta,\tilde{\xi} )  \right]^2 } }{1 + \rho \left[ \tau^2 + \frac{1-\tau^2}{\left(1 + g ( \beta,\tilde{\xi} )\right)^2} \right]} \right]^+.
\label{eqn:Rs_imperfect_CSI}
\end{equation}
\end{Theorem}
\begin{IEEEproof}
The proof of Theorem 4 is provided in Appendix B.
\end{IEEEproof}

\subsection{Minimum Required CSI for the RCI-PR Precoder}

We now consider our proposed RCI-PR precoder, and determine how the CSI estimation error must scale with the SNR, to maintain a given rate gap to the case with perfect CSI. In the following, we assume that the regularization parameter $\xi^{\star\circ}$ from (\ref{eqn:xi_opt}) is used. This does not require the transmitter to be aware of the value of the distortion $\tau^2$. We define the per-user gap $\Delta R_s^{\circ}$ as the difference 
\begin{equation}
\Delta R_s^{\circ} \triangleq \frac{R_{s,\mathrm{PR}}^{\star\circ}-\hat{R}_{s,\mathrm{PR}}^{\star\circ}}{K}
\end{equation}
where $R_{s,\mathrm{PR}}^{\star\circ}$ and $\hat{R}_{s,\mathrm{PR}}^{\star\circ}$ are the large-system secrecy sum-rates obtained by the RCI-PR precoder under perfect CSI and under CSI distortion $\tau^2$, respectively. We now derive the scaling of $\tau^2$ required to maintain a constant secrecy rate gap for high SNR, so that the multiplexing gain is not affected.

\begin{Proposition}
For $\beta \leq 1$, $b>1$, a CSI distortion $\tau^2 = \frac{C}{\rho}$, with
\begin{equation}
C = \left\{ 
  \begin{array}{l l l}
     \frac{1}{2} \left( \sqrt{4b-3} - 1 \right) & \quad \textrm{for $\beta < 1$} \\
     \frac{2}{3} \left( \sqrt{3b-2} - 1 \right) & \quad \text{for $\beta = 1$} \\
  \end{array} \right.
\label{eqn:C}
\end{equation}
produces a high-SNR rate gap of $\log_2 b $ bits.
\label{prop:tau_1}
\end{Proposition}
\begin{IEEEproof}
For $\beta \leq 1$, define
\begin{equation}
\mu \triangleq \tau^2\rho+\frac{\tau^4\rho(1+g ( \beta,\xi^{\star\circ} ))^2}{1-\tau^2}+\frac{\tau^2(1+g ( \beta,\xi^{\star\circ} ))^2}{1-\tau^2}.
\end{equation}
We have
\begin{equation}
\lim\limits_{\rho \rightarrow \infty} \Delta R_s^{\circ} = \left\{ 
  \begin{array}{l l l}
    \lim\limits_{\rho \rightarrow \infty} \log_2 \left[ 1+\frac{\beta^2}{4\rho(1-\beta)^2}  \mu\right] = \log_2 b & \quad \textrm{for $\beta < 1$} \\
   \lim\limits_{\rho \rightarrow \infty} \log_2 \left[  1+\frac{1}{4} \mu\right] = \log_2 b & \quad \text{for $\beta = 1$} \\
  \end{array} \right.
\label{eqn:proof_prop}
\end{equation}
\end{IEEEproof}

\begin{Proposition}
For $\beta > 1$, if $\lim_{\rho \rightarrow \infty} \tau^2 = 0$, then the high-SNR rate gap is zero.
\label{prop:tau_2}
\end{Proposition}
\begin{IEEEproof}
For $\beta > 1$, defining
\begin{equation}
\nu \triangleq \left[\tau^2\rho^{\star\circ}+\frac{\tau^4\rho^{\star\circ}\beta^2}{\left(\beta-1\right)^2\left(1-\tau^2\right)}+\frac{\tau^2\beta^2}{\left(\beta-1\right)^2\left(1-\tau^2\right)}\right]
\end{equation}
we have
\begin{equation}
\lim_{\rho \rightarrow \infty} \Delta R_s^{\circ} = \lim_{\rho \rightarrow \infty} \log_2 \left[1 + \left(1-\frac{\beta}{2}\right)\nu\right] = 0.
\end{equation}
\end{IEEEproof}

We now consider the case of FDD systems. We assume that users quantize their channel directions by using $B$ bits and employing random vector quantization (RVQ), and that they feed the quantization index back to the transmitter \cite{Jindal06,Ryan09}. We obtain the following result.
\begin{Proposition}
In order to maintain a high-SNR secrecy rate offset of $\log_2 b$ bits per-user in the large-system regime and for all values of the network load $\beta$, it is sufficient to scale the number of feedback bits $B$ per user as $B \approx \frac{M-1}{3} \rho_{\textrm{dB}} - \left( M-1 \right) \left[ \log_2 \left( \sqrt{4b-3} - 1 \right)-1\right]$.
\label{prop:feedback_bits}
\end{Proposition}
\begin{IEEEproof}
We note from Propositions \ref{prop:tau_1} and \ref{prop:tau_2} that $\tau^2 =\frac{C}{\rho}$ ensures a gap of $\log_2 b$ bits $\forall \beta$. If RVQ is used, then the quantization error $\tau^2$ can be upper bounded as $\tau^2 < 2^{-\frac{B}{M-1}}$ \cite{Jindal06}. Therefore, it is sufficient to scale the number of feedback bits per user according to $B = \left( M-1 \right) \log_2 \rho - \left( M-1 \right) \log_2 C$. Substituting the smallest value of $C$ from (\ref{eqn:C}) and rewriting $\rho$ in dB yields Proposition \ref{prop:feedback_bits}.
\end{IEEEproof}

\subsection{Numerical Results}

\begin{figure}
\centering
\includegraphics[width=\columnwidth]{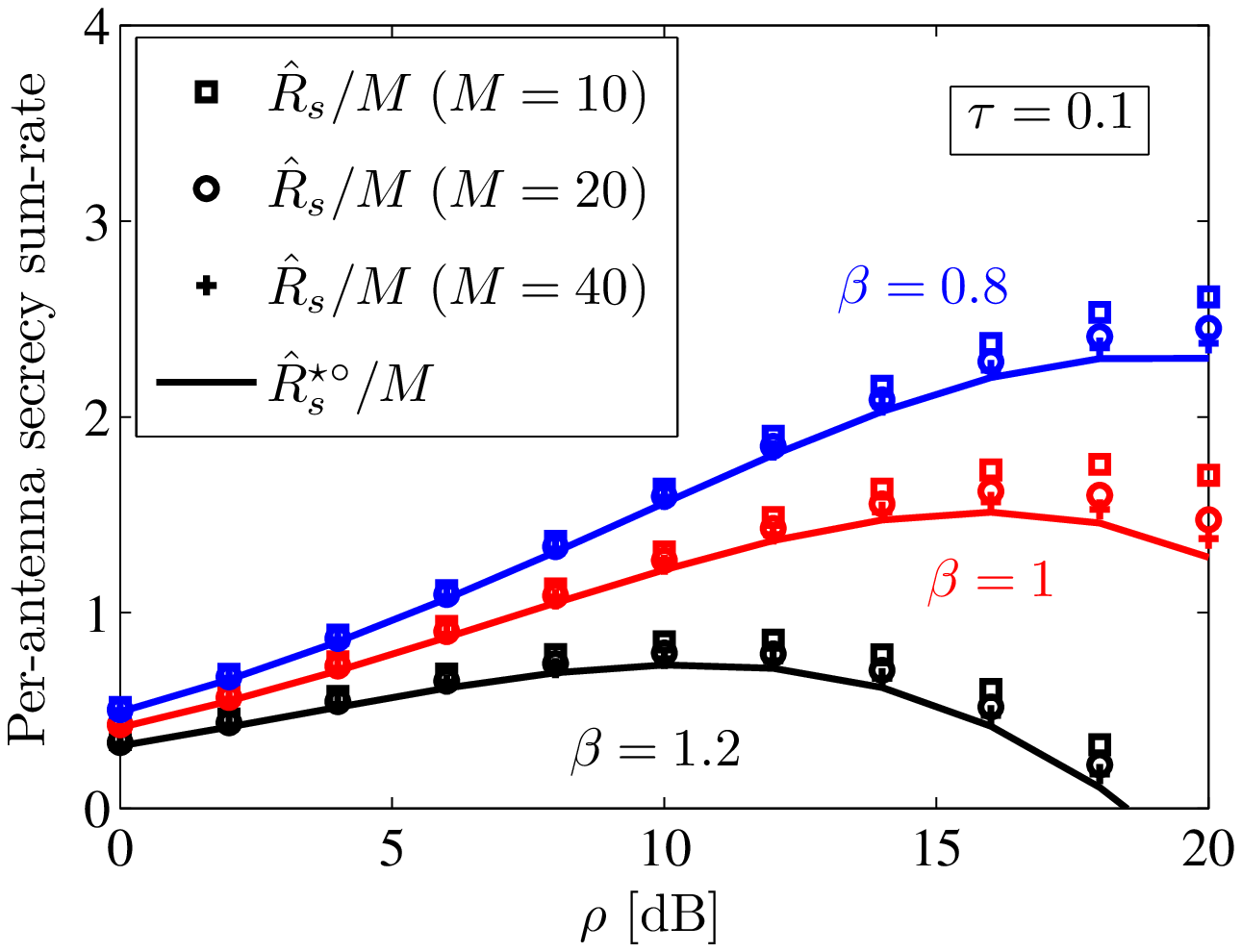}
\caption{Comparison between the per-antenna secrecy sum-rate $\hat{R}_s^{\star\circ}/M$ with RCI precoding in the large-system regime (\ref{eqn:Rs_imperfect_CSI}) and the simulated ergodic secrecy sum-rate $\hat{R}_s/M$, for finite $M$, and in the presence of a channel estimation error $\tau=0.1$. Three sets of curves are shown, each one corresponds to a different value of $\beta$.}
\label{fig:Noisy_CSI_20dB}
\end{figure}

Fig. \ref{fig:Noisy_CSI_20dB} compares the secrecy sum-rate $\hat{R}_s^{\star\circ}$ of the RCI precoder from the large-system analysis to the simulated ergodic secrecy sum-rate $\hat{R}_s$ for finite $M$, in the presence of a CSI error $\tau=0.1$ and for different values of $\beta$. The values of $\hat{R}_s^{\star\circ}$ and $\hat{R}_s$ were obtained by (\ref{eqn:Rs_imperfect_CSI}) and (\ref{eqn:Rs_imperfect_CSI_finite}), respectively, with $\xi = \xi^{\star\circ}$. As expected, the accuracy of the deterministic equivalent increases as $M$ grows.

\begin{figure}
\centering
\includegraphics[width=\columnwidth]{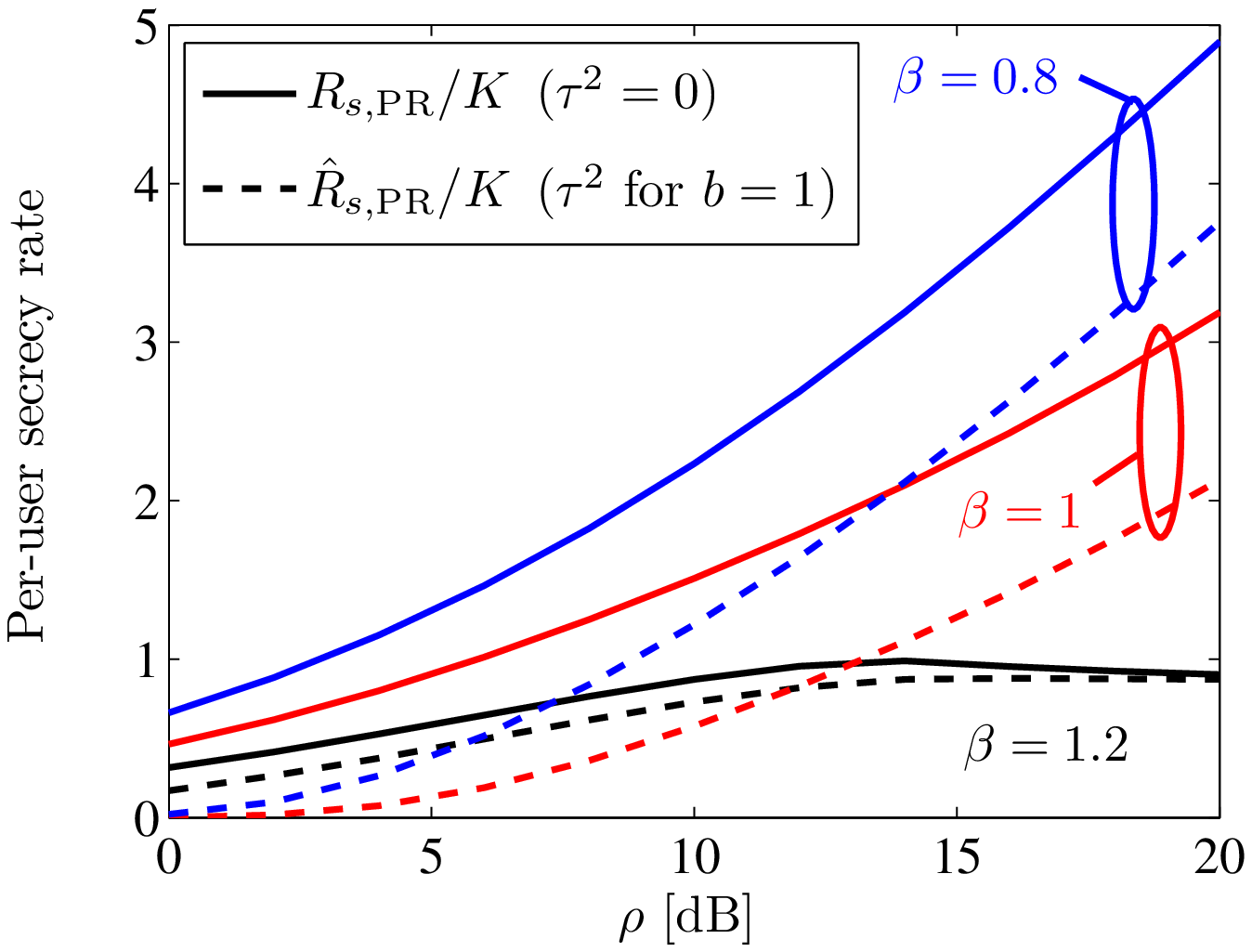}
\caption{Comparison between the ergodic per-user secrecy rates $R_{s,\mathrm{PR}}$ and $\hat{R}_{s,\mathrm{PR}}$ with RCI-PR precoder in the presence of perfect CSI and in the presence of channel estimation error $\tau^2$, respectively. For $\beta>1$, we used $\tau^2=\frac{0.1}{\rho}$. For $\beta\leq1$, we used $\tau^2$ from Proposition \ref{prop:tau_1}, with $\log_2 b = 1$ bit. Three sets of curves are shown for $M=10$, each one corresponds to a different value of $\beta$.}
\label{fig:Limited_feedback}
\end{figure}

Fig. \ref{fig:Limited_feedback} shows the ergodic per-user secrecy rate $\hat{R}_{s,\mathrm{PR}}/K$, achieved by the proposed RCI-PR precoder in the presence of a channel estimation error that scales: (i) as $\tau^2=\frac{0.1}{\rho}$ for $\beta>1$, and (ii) as in Proposition \ref{prop:tau_1} for $\beta\leq1$, with $\log_2 b = 1$ bit. This is compared to the ergodic rate $R_{s,\mathrm{PR}}/K$, achieved by the same precoder in the presence of perfect CSI ($\tau=0$). Three different values of $\beta$ are considered, and $M=10$. For $\beta \leq 1$, the simulations show a high-SNR gap of nearly $1$ bit, whereas for $\beta>1$ no gap is present. These results confirm the claims made in Propositions \ref{prop:tau_1} and \ref{prop:tau_2}.
\section{Conclusion}
In this paper, we considered linear precoding for the MISO broadcast channel with confidential messages. We first studied the RCI precoder in the large-system regime, and obtained a deterministic equivalent for the achievable secrecy sum-rate, as well as expressions for the optimal regularization parameter $\xi$ and network load $\beta$. This analysis proved to be accurate even for finite-size systems. We found that for $\beta>1$ the RCI precoder performs poorly in the high-SNR regime. We therefore proposed a linear precoder based on RCI and power reduction (RCI-PR) to increase the high-SNR performance for network loads in the range $1 < \beta < 2$. The proposed RCI-PR precoder was showed to achieve a per-user secrecy rate with the same high-SNR scaling factor as both the following upper bounds: (i) the sum-rate of the optimal RCI precoder in the absence of secrecy requirements, and (ii) the secrecy capacity of a single-user system without interference. We further studied the case of imperfect CSI, and we derived a deterministic equivalent for the secrecy sum-rate achievable by RCI precoding. We finally considered the performance of our proposed RCI-PR precoder in the presence of CSI error, and determined: (i) how the CSI estimation error must scale with the SNR, and (ii) how many feedback bits are required, in order to maintain a given rate gap to the case with perfect CSI.

The authors are currently working to study the effect of limited feedback on time division duplex (TDD) systems, deriving the secrecy sum-rate as a function of the amount of channel training, as well as the optimal amount of channel training that maximizes the secrecy sum-rate \cite{GeraciICASSP13}. Moreover, the authors are working to generalize the results present in this paper to the case where channel correlation is present at the transmitter side, and to study how correlation affects the secrecy rates achievable by RCI precoding \cite{GeraciLetter}. 
\appendices
\section{Proof of Theorem 1}

By defining
\begin{equation}
A_k = \mathbf{h}_k^H \left( \mathbf{H}_k^H \mathbf{H}_k + M \xi \mathbf{I} \right) ^{-1} \mathbf{h}_k,
\label{eqn:Ak}
\end{equation}
\begin{equation}
B_k = \mathbf{h}_k^H \left(\mathbf{H}_k^H \mathbf{H}_k + M \xi \mathbf{I} \right) ^{-1}  \mathbf{H}_k^H \mathbf{H}_k \left(  \mathbf{H}_k^H \mathbf{H}_k + M \xi \mathbf{I} \right) ^{-1}  \mathbf{h}_k
\label{eqn:Bk}
\end{equation}
we can rewrite the SINRs at the intended user and the eavesdropper as
\begin{equation}
\mathrm{SINR}_{k} = \frac{\rho A_k^2}{\gamma \left( 1 + A_k \right)^2 + \rho B_k} \quad \textrm{and} \quad
\mathrm{SINR}_{\widetilde{k}} = \frac{\rho B_k}{\gamma \left( 1 + A_k \right)^2 }.
\label{eqn:SINR_rewritten}
\end{equation}

We rewrite $A_k = \frac{M}{K-1} \mathbf{v}^H_M \mathbf{Q}_M \mathbf{v}_M$, where
\begin{equation}
\mathbf{Q}_{k,M} = \left( \frac{1}{K-1}\mathbf{H}^H_k\mathbf{H}_k  - z \mathbf{I} \right)^{-1} \in \mathbb{C}^{M \times M}, \quad \mathbf{v}_{k,M} = \frac{1}{\sqrt{M}} \mathbf{h}_k \in \mathbb{C}^{M}, \quad z = -\frac{M \xi}{K-1}.
\end{equation}
By Bai and Silverstein's Lemma \cite{Peacock08,CouilletBook,Couillet12Robust}, we have
\begin{equation}
\textrm{E} \left[ \phi_k \left| \mathbf{v}_{k,M}^H \mathbf{Q}_{k,M} \mathbf{v}_{k,M} - \frac{1}{M} \textrm{tr} \mathbf{Q}_{k,M} \right|^p \right] \leq \frac{C_p}{M^p}  \left(\frac{M}{\bar{\lambda}_M}\right)^{\frac{p}{2}} = \frac{C_p}{M^{\frac{p}{2}}\bar{\lambda}_M^{\frac{p}{2}}} = f_M \quad \forall p \geq 1
\end{equation}
where $C_p$ is a constant depending only on $p$, $\phi_k = 1_{\{ |\lambda_1(\mathbf{H}_k^H\mathbf{H}_k)-M\xi|,\ldots,|\lambda_M(\mathbf{H}_k^H\mathbf{H}_k)-M\xi|>\bar{\lambda}_M\}}$, with $\bar{\lambda}_M\rightarrow0$. Assume $\xi \in \mathcal{D}'_M$, with $\mathcal{D}'_M = \mathcal{D}_M$ for $\beta \leq 1$ and $\mathcal{D}'_M = \mathcal{D}_M \backslash \left\{ \left[ - \frac{C}{M^{\frac{1}{2}-\epsilon}}, + \frac{C}{M^{\frac{1}{2}-\epsilon}} \right] \right\}$ for $\beta>1$. Then we have $M = \mathcal{O}(\bar{\lambda}_M^{-2-\epsilon})$, for some $\epsilon > 0$, and $\underset{k \leq K}{\min}\{\phi_k\}\stackrel{\textrm{a.s.}}{\longrightarrow}1$ \cite{Couillet12Robust}. It follows from the Markov inequality and the Borel-Cantelli lemma \cite{BillingsleyBook} that $\underset{k}{\max}|\mathbf{v}_{k,M}^H \mathbf{Q}_{k,M} \mathbf{v}_{k,M} - \frac{1}{M} \textrm{tr} \mathbf{Q}_{k,M}| \stackrel{\textrm{a.s.}}{\longrightarrow} 0$, as $M \rightarrow \infty$.
The term $\frac{1}{M} \textrm{tr} \mathbf{Q}_{k,M}$ is by definition the Stieltjes transform $m_{\mathbf{H}^H_k\mathbf{H}_{k,M}}(z)$ of $\mathbf{H}^H_k\mathbf{H}_{k,M}$. Similarly, it can be shown that $\underset{k}{\max}|m_{\mathbf{H}^H_k\mathbf{H}_{k,M}}(z) - m(z)|\stackrel{\textrm{a.s.}}{\longrightarrow} 0$,
where $m(z)$ can be obtained as the solution of $m(z) = \left[1 - \frac{M}{K-1} - z -z \frac{M}{K-1} m(z)\right]^{-1}$.
This yields
\begin{equation}
A_k - g \left( \beta,\xi \right) \stackrel{\textrm{a.s.}}{\rightarrow} 0
\label{eqn:Ak_as}
\end{equation}
with
\begin{equation}
g \left( \beta,\xi \right) = \beta^{-1} m(\xi) = \frac{1}{2} \left[ \pm \sqrt{ \frac{\left(1-\beta \right)^2}{\xi^2}  +  \frac{2\left(1+\beta\right)}{\xi} +  1} +  \frac{1-\beta}{\xi}  -  1 \right],
\end{equation}
and where in order for $m$ to be a Stieltjes transform, the sign of the square root must be chosen the same as the sign of $\xi$ \cite{SilversteinChoi}.

We now rewrite $B_k = A_k - \frac{M^2 \xi}{\left(K-1\right)^2} \mathbf{v}_{k,M}^H \mathbf{Q}_{k,M}^2 \mathbf{v}_{k,M}$,
and similarly we have
\begin{equation}
\textrm{E} \left[ \phi_k \left| \mathbf{v}_{k,M}^H \mathbf{Q}^2_{k,M} \mathbf{v}_{k,M} - \frac{1}{M} \textrm{tr} \mathbf{Q}^2_{k,M} \right|^p \right] \leq \frac{C_p}{M^p}  \left(\frac{M}{\bar{\lambda}_M^2}\right)^{\frac{p}{2}} = \frac{C_p}{M^{\frac{p}{2}}\bar{\lambda}_M^p} = g_M \quad \forall p \geq 1.
\end{equation}
Again, if $M = \mathcal{O}(\bar{\lambda}_M^{-2-\epsilon})$, for some $\epsilon > 0$, we have $\underset{k}{\max}|\mathbf{v}_{k,M}^H \mathbf{Q}^2_{k,M} \mathbf{v}_{k,M} - \frac{1}{M} \textrm{tr} \mathbf{Q}^2_{k,M}| \stackrel{\textrm{a.s.}}{\longrightarrow} 0$, as $M \rightarrow \infty$.
We note that $\frac{1}{M} \textrm{tr} \mathbf{Q}^2_{k,M}$ is the Stieltjes transform of $\mathbf{Q}^2_{k,M}$, given by
\begin{equation}
\frac{1}{M} \textrm{tr} \mathbf{Q}^2_{k,M} = \int {\frac{dF_M(\lambda)}{\left(\lambda_{\mathbf{R}_M} - z\right)^2}} = \frac{\partial}{\partial z} \int {\frac{dF_M(\lambda)}{\lambda_{\mathbf{R}_M} - z}} = m'_{\mathbf{R}_M}(z)
\end{equation}
where $F_M(\lambda)$ is the distribution of the eigenvalues of $\mathbf{R}_M$. Since both $m_{\mathbf{R}_{k,M}}(z)$ and $m(z)$ are analytic functions, we have $\underset{k}{\max}|m'_{\mathbf{R}_{k,M}}(z) - m'(z)| \stackrel{\textrm{a.s.}}{\longrightarrow} 0$, as $M \rightarrow \infty
$,
and it follows that
\begin{equation}
B_k - \left[ \frac{1}{\beta} m(z) - \frac{M^2 \xi}{\left(K-1\right)^2} \frac{\partial}{\partial z} m(z) \right] = B_k - \left[ g \left( \beta,\xi \right) + \xi \frac{\partial}{\partial \xi} g \left( \beta,\xi \right) \right] \stackrel{\textrm{a.s.}}{\longrightarrow} 0.
\label{eqn:Bk_as}
\end{equation}

For the power normalization constant $\gamma$ we have
\begin{equation}
\gamma = \mathrm{tr} \left\{ \left( \mathbf{H}^H \mathbf{H} + M \xi \mathbf{I} \right) ^{-1} \right\} - M \xi \mathrm{tr} \left\{ \left( \mathbf{H}^H \mathbf{H} + M \xi \mathbf{I} \right) ^{-2} \right\} = \frac{1}{\beta} m_{\mathbf{H}\mathbf{H}^H_M}(z') - \frac{\xi}{\beta^2} m'_{\mathbf{H}\mathbf{H}^H_M}(z')
\end{equation}
where $z' = -\frac{M \xi}{K}$.
If $\xi \in \mathcal{D}'_M$, if follows that
\begin{equation}
\gamma - \left[ g \left( \beta,\xi \right) + \xi \frac{\partial}{\partial \xi} g \left( \beta,\xi \right) \right] \stackrel{\textrm{a.s.}}{\longrightarrow} 0.
\label{eqn:gamma_as}
\end{equation}
By the continuity of $R_s$ and $R_s^{\circ}$, it follows that the previous convergence results also hold for $\xi \in \left[ - \frac{C}{M^{\frac{1}{2}-\epsilon}}, + \frac{C}{M^{\frac{1}{2}-\epsilon}} \right]$ and $\beta > 1$. Equation (\ref{eqn:Theorem1}) then follows from (\ref{eqn:Rs}), (\ref{eqn:SINR_rewritten}), (\ref{eqn:Ak_as}), (\ref{eqn:Bk_as}), (\ref{eqn:gamma_as}), and by applying the continuous mapping theorem, the Markov inequality, and the Borel-Cantelli lemma.
\section{Proof of Theorem 4}
From \cite{Nguyen09}, by defining $\tilde{\rho} \stackrel{\triangle}{=} \frac{\rho\left(1-\tau^2\right)}{\rho \tau^2 + 1}$ and $\tilde{\xi} \stackrel{\triangle}{=} \frac{\xi}{1 - \tau^2}$, a deterministic equivalent for $\hat{\textrm{SINR}}_k$ is given by
\begin{equation}
\hat{\textrm{SINR}}_k^{\circ} = g ( \beta,\tilde{\xi} ) \frac{\tilde{\rho} + \frac{ \tilde{\xi} \tilde{\rho}}{\beta} \left[ 1 + g ( \beta,\tilde{\xi} ) \right]^2}{\tilde{\rho} + \left[ 1 + g ( \beta,\tilde{\xi} )  \right]^2 }.
\label{eqn:SINR_hat_deteq}
\end{equation}
By defining $\mathbf{\Omega}_k = \left( \hat{\mathbf{H}}_k^H \hat{\mathbf{H}}_k + M \xi \mathbf{I} \right) ^{-1}
$, we can rewrite $\hat{\textrm{SINR}}_{\tilde{k}}$ as
\begin{equation}
\hat{\textrm{SINR}}_{\tilde{k}} = \rho \frac{\hat{B}_k+2\left(1+\hat{A}_k\right)Q_k+\left(1+\hat{A}_k\right)^2R_k}{\hat{\gamma}\left( 1+\hat{A}_k \right)^2},
\end{equation}
where
\begin{equation*}
\hat{A}_k = \hat{\mathbf{h}}_k^H \mathbf{\Omega}_k \hat{\mathbf{h}}_k, \quad \hat{B}_k = \hat{\mathbf{h}}_k^H \mathbf{\Omega}_k \hat{\mathbf{H}}_k^H \hat{\mathbf{H}}_k \mathbf{\Omega}_k \hat{\mathbf{h}}_k,
\end{equation*}
\begin{equation*}
Q_k = \hat{\mathbf{h}}_k^H \mathbf{\Omega}_k \hat{\mathbf{H}}_k^H \hat{\mathbf{H}}_k  \left( \mathbf{\Omega}_k - \frac{\mathbf{\Omega}_k \hat{\mathbf{h}}_k \hat{\mathbf{h}}_k^H \mathbf{\Omega}_k}{1+\hat{A}_k} \right) \mathbf{e}_k, \quad \textrm{and}
\end{equation*}
\begin{equation*}
R_k = \mathbf{e}_k^H \left( \mathbf{\Omega}_k - \frac{\mathbf{\Omega}_k \hat{\mathbf{h}}_k \hat{\mathbf{h}}_k^H \mathbf{\Omega}_k}{1+\hat{A}_k} \right) \hat{\mathbf{H}}_k^H \hat{\mathbf{H}}_k  \left( \mathbf{\Omega}_k - \frac{\mathbf{\Omega}_k \hat{\mathbf{h}}_k \hat{\mathbf{h}}_k^H \mathbf{\Omega}_k}{1+\hat{A}_k} \right) \mathbf{e}_k.
\end{equation*}
If $\xi \in \mathcal{D}_M$, we have
\begin{equation*}
\hat{A}_k - g \left( \beta,\tilde{\xi} \right)\stackrel{\textrm{a.s.}}{\longrightarrow} 0, \quad \hat{B}_k  - \left[ g \left( \beta,\tilde{\xi} \right) + \tilde{\xi} \frac{\partial}{\partial \tilde{\xi}} g \left( \beta,\tilde{\xi} \right) \right] \stackrel{\textrm{a.s.}}{\longrightarrow} 0, \quad Q_k \stackrel{\textrm{a.s.}}{\longrightarrow} 0,
\end{equation*}
\begin{equation*}
R_k - \frac{\tau^2}{1-\tau^2}\left[ g \left( \beta,\tilde{\xi} \right) + \tilde{\xi} \frac{\partial}{\partial \tilde{\xi}} g \left( \beta,\tilde{\xi} \right) \right] \stackrel{\textrm{a.s.}}{\longrightarrow} 0, \quad \textrm{and} \quad \hat{\gamma} - \frac{1}{1-\tau^2}\left[ g \left( \beta,\tilde{\xi} \right) + \tilde{\xi} \frac{\partial}{\partial \tilde{\xi}} g \left( \beta,\tilde{\xi} \right) \right] \stackrel{\textrm{a.s.}}{\longrightarrow} 0
\end{equation*}
hence a deterministic equivalent for $\hat{\textrm{SINR}}_{\tilde{k}}$ is given by
\begin{equation}
\hat{\textrm{SINR}}_{\tilde{k}}^{\circ} = \rho \left[ \tau^2 + \frac{1-\tau^2}{\left(1 + g ( \beta,\tilde{\xi} )\right)^2} \right].
\label{eqn:SINR_tilde_hat_deteq}
\end{equation}
Theorem 4 then follows from (\ref{eqn:Rs_imperfect_CSI_finite}), (\ref{eqn:SINR_hat_deteq}), (\ref{eqn:SINR_tilde_hat_deteq}), and from the continuous mapping theorem \cite{BillingsleyBook}.
\ifCLASSOPTIONcaptionsoff
  \newpage
\fi
\bibliographystyle{IEEEtran}
\bibliography{IEEEabrv,Bib_Giovanni}
\end{document}